\documentclass[reprint,
groupedaddress,
amsmath,amssymb,floatfix,
prb,
]{revtex4-2}

\usepackage{xr-hyper}
\usepackage{hyperref}
\usepackage[utf8]{inputenc}
\usepackage{graphicx}
\usepackage{dcolumn}
\usepackage{bm}
\usepackage[mathlines]{lineno}
\usepackage{xcolor}
\usepackage{soul}
\usepackage{lineno}

\newcommand{\E}{\mathcal{E}}
\newcommand{\q}{\mathbf{q}}
\makeatletter

\newcommand*{\addFileDependency}[1]{
  \typeout{(#1)}
  \@addtofilelist{#1}
  \IfFileExists{#1}{}{\typeout{No file #1.}}
}
\makeatother

\newcommand*{\myexternaldocument}[1]{%
    \externaldocument{#1}%
    \addFileDependency{#1.tex}%
    \addFileDependency{#1.aux}%
}

\myexternaldocument{SI_v3}

\begin{document}
\title{Theory of non-equilibrium ``hot'' carriers in direct band-gap semiconductors under continuous illumination}

\author{Subhajit Sarkar}
\email{subhajit@post.bgu.ac.il}
\affiliation{Department of Chemistry, Ben-Gurion University of the Negev}
\affiliation{School of Electrical and Computer Engineering, Ben-Gurion University of the Negev}

\author{Ieng-Wai Un and Yonatan Sivan}
\affiliation{School of Electrical and Computer Engineering, Ben-Gurion University of the Negev}

\author{Yonatan Dubi}
\email{jdubi@bgu.ac.il}
\affiliation{Department of Chemistry, Ben-Gurion University of the Negev
}
\affiliation{Ilse Katz Center for Nanoscale Science and Technology, Ben-Gurion University of the Negev
}

\date{\today}

\begin{abstract}
The interplay between the illuminated excitation of carriers and subsequent thermalization and recombination leads to the formation of non-equilibrium distributions for the ``hot'' carriers and to heating of both electrons, holes and phonons. In spite of the fundamental and practical importance of these processes, there is no theoretical framework which encompasses all of them and provides a clear prediction for the non-equilibrium carrier distributions. Here, a self-consistent theory accounting for the interplay between excitation, thermalization, and recombination in continuously-illuminated semiconductors is presented, enabling the calculation of non-equilibrium carrier distributions. We show that counter-intuitively, distributions deviate more from equilibrium under weak illumination than at high intensities. We mimic two experimental procedures to extract the carrier temperatures and show that they yield different dependence on illumination. Finally, we provide an accurate way to evaluate photoluminescence efficiency, which, unlike conventional models, predicts correctly the experimental results. These results provide a starting point towards examining how non-equilibrium features will affect properties hot-carrier based application. 
\end{abstract}

\maketitle
\section{Introduction} Illumination of semiconductors causes the generation of high energy non-thermal (aka ``hot'') carriers (HCs)~\cite{SHAH19923} which critically impacts the performance of semiconductor-based electronic and opto-electronic devices~\cite{Takeda-Suzuki1983,Green2017,Nozik2018}. Although the ``hot'' carrier (HC) dynamics in semiconductors have been studied for the past few decades~\cite{SHAH19923,Takeda-Suzuki1983}, the topic has seen a growing interest in recent times due to technological relevance~\cite{Green2017,Konig20102862,CONIBEER2009713,RevModPhys-Rossi-Kuhn,Sjakste_2018JPCM}. Much experimental and theoretical efforts have been devoted to studying the {\em transient} dynamics of HCs in semiconductors under {\em ultrafast} pulsed illumination~\cite{Shah-ch4,elsaesser1991initial,schoenlein1987femtosecond,Doany-1988,Bernardi5291,SelbmannPhysRevB1996,Shah-ch3, RevModPhys-Rossi-Kuhn, fischetti1988monte,jacoboni1983monte,SnokePRB1992,SjakstePhysRevB2018,Sjakste_2018JPCM, Wittenbecher_acsnano_2021} with the aim of understanding the HC thermalization time scales and role of HC-phonon interaction that can affect performance of HC based solar energy conversion \cite{ferry2019search, esmaielpour2020exploiting} and lighting applications (e.g., photo-detection, HC based lasers) \cite{Lao2014, Elsasser_HC_laser}. 

Various key applications of semiconductors, such as solar-cells~\cite{Green2017,Konig20102862,CONIBEER2009713,Polman2012,HirstAPL2014,LombezAPL2015,GradauskasAPL2018} and lighting applications~\cite{LED-book,Li2006Semiconductor-Physical-Electronics} involve instead {\em continuous-wave} (CW) illumination, under which the system is necessarily in a non-equilibrium {\em steady-state} (NESS). The NESS consists of a carrier distribution that deviates from the thermal (i.e., Fermi-Dirac) distribution, and may also involve carrier temperatures differing from each other and from the lattice temperature. 

Despite the extensive work on the topic, most studies involved only a macroscopic description of the problem, while the fewer existing state-of-the-art modelling of the NESS have several limitations. First, the tenet that the photo-excited HCs ultimately achieve the lattice temperature in the steady-state~\cite{Nozik_annurev.physchem} has been challenged in a recent experiment on group III-V semiconductors; it unveils that in the NESS carrier temperatures can be much higher than lattice temperature~\cite{Tedeschi_acs.nanolett_2016, Shojaei2019, Chen2020, Fast_2020}. A theoretical understanding of the existence of such a high carrier-temperature in NESS is still lacking~\cite{Tedeschi_acs.nanolett_2016, Shojaei2019, Chen2020, Fast_2020}, especially in light of the different techniques available for measuring the various temperatures. 
Second, the existing theoretical studies of the steady-state HC dynamics have considered the HC relaxation and interband recombination within the relaxation time approximation (RTA) where the scattering rate is assumed to be independent of the NESS distributions~\cite{Kamide_PhysRevApplied,Takeda_2010,Kamide_JAPL,K_nig_2020}. 
Although the RTA makes the calculation of NESS distribution simpler, it neither conserves particle number nor energy within each band. Moreover, a constant (distribution-independent) relaxation rate may overestimate/underestimate the scattering rates, such that consequently, the results become inaccurate.
In addition, previous studies neglected the heating of the lattice~\cite{Dimmock2014,Tsai2018,Tsai2019}, thus, they cannot treat correctly the amount of energy dissipated to the environment from the lattice, overall making it impossible to assess the difference between the carrier, phonon and environment temperatures. 
Last but not least, theoretical modelling of semiconductor photoluminescence (PL) and cooling of SCs has been based on carrier-only (macroscopic) rate equations of the total number density (rather than the Fermi golden rule expression which relies on the energy distributions of electrons and holes)~\cite{Hall-SRH-1952,Shockley-Read-SRH-1952,Johnson_JVST_2007_Auger,chuang2009physics,Li2006Semiconductor-Physical-Electronics, Andre-20, mork1992chaos}, thus, neglecting the role of carrier-phonon interactions and system-environment coupling~\cite{Hess-Kuhn-PRA-1996,Hannewald-PhysRev-PL-2000,Hannewald-PhysRevB.67.233202,PhysRevLett-LaserCooling-2004}; this, again, leads to an inaccurate assessment of the lattice temperatures and even led to some debate in temperature measurements in optical cooling experiments~\cite{Morozov2019}. 
The above list of approximations indicate the lack of a comprehensive theoretical approach to understand the NESS properties of semiconductors under CW illumination that incorporates full non-equilibrium distributions in determining carrier excitation, recombination, carrier-phonon and carrier-carrier scatterings. 

\textcolor{black}{ Motivated by the semiconductor applications perspective, the lack of a complete theory incorporating the non-equilibrium nature of the photo-excited carriers, and the advance in the theoretical formulation in metallic systems~\cite{Dubi2019, jermyn2019transport, lozan2017increased, PhysRevLett_Brown_2017}, we put forward a theoretical formalism for the non-equilibrium dynamics of HCs under CW illumination.} Our primary aim is to describe the steady-state properties. In particular, we present a set of coupled Boltzmann-heat equations for the non-equilibrium photo-generated carrier distributions, whereby photo-excitation, carrier-carrier and carrier-phonon scattering rates, and recombination of electrons and holes are evaluated self-consistently within the Fermi golden rule, thus introducing (semi-) quantum behavior into the Boltzmann equation (BE). The macroscopic properties (e.g., the power flows and the phonon temperature) are obtained by integrating over the distributions and employing the conservation of particle number and energy for each subsystem~\cite{Dimmock2014,Tsai2018,Tsai2019}. 



Incorporation of the full non-equilibrium distributions in a self-consistent way along with the energy and number conservation distinguishes our work from existing formalisms that treat the non-equilibrium \textcolor{black}{in semiconductors} only within RTA~\cite{Dimmock2014,Tsai2018,Tsai2019,Kamide_PhysRevApplied,Takeda_2010,Kamide_JAPL,K_nig_2020}. 
Our central results can be summarized as follows: 
\begin{itemize}
    \item Counter-intuitively, we find the NESS carrier distribution is more out-of-equilibrium (i.e., less resembles an equilibrium distribution), indicating higher density of non-thermal carriers at low illumination intensities than at high intensities~\cite{NEGES20062107} due to inefficient thermalization at low intensities. An equivalent interpretation commonly adopted in semiconductor textbooks \cite{Ashcroft-Mermin, SzeCh1} is to interpret the distribution as being thermal with yet a higher effective chemical potential. 

    \item Consequently, the temperatures of the carriers become ill-defined at low illumination, and depend on the way the temperatures are measured. We show this by extracting carrier temperatures using two generic experimental procedures frequently used, namely (i) extracting a temperature from the heat transfer between carriers and phonons (measured via, e.g., a floating thermal probe or a thermocouple~\cite{dubi2011colloquium,cui2017study}), and (ii) by fitting the photoluminescence (PL) spectra~\cite{HirstInGaAs-PL,Nguyen2018_Nature_Energy,Giteau-JAP2020}. These two temperatures are shown to be very different from one another. 

\item Finally, our formalism allows us to evaluate the energy partition for an illuminated semiconductor, i.e., how much of the power pumped into the system goes to heating-up the lattice, and how much remains in the ``hot" electron sub-system and subsequently (partially) dissipates through radiative recombination. 
We show that this partition depends on the energy of the single photon but not on the total number of photons, and explain experimental observations of the PL efficiency in CW-illuminated Galium Arsenide (GaAs). 
\end{itemize}

The paper is organized as follows. in Sec. \ref{formulation} we describe our microscopic formulation \textcolor{black}{and clearly state the simplifying assumptions as a first step, e.g., we consider parabolic band structure. We also describe why the assumption made are well suited in the context of many III-V semiconductors, in particular, for GaAs.} Our formulation of photo-excitation and recombination are distinctive due to the fact, unlike the existing formulations, e.g., Refs. \cite{Dimmock2014,Tsai2018,Tsai2019,Kamide_PhysRevApplied,Takeda_2010,Kamide_JAPL,K_nig_2020}, ours incorporate the full non-equilibrium distribution explicitly. \textcolor{black}{Our formulation of carrier-phonon interaction, although incorporating acoustic phonons only, can easily be extended to other phonon modes such as optical phonons or even piezoelectric phonons \cite{ridley1999quantum}. We emphasize that a comparative study of the role of different phonon modes in the steady-state properties is beyond the scope of the present paper. We incorporate carrier-carrier interaction within hard-sphere approximation using Fermi golden rule which is a good approximation in most of the undoped III-V semiconductors due to low carrier densities.} In Sec. \ref{sec:macro} we describe the macroscopic quantities that are essential for defining a steady-state. In Sec. \ref{sec:results} we describe our results, and conclude and indicate possible improvements to our formalism in Sec \ref{sec: conclusion}. In Appendices we describe the necessary mathematical details of our formulation.
\begin{figure}
\includegraphics[scale=0.35]{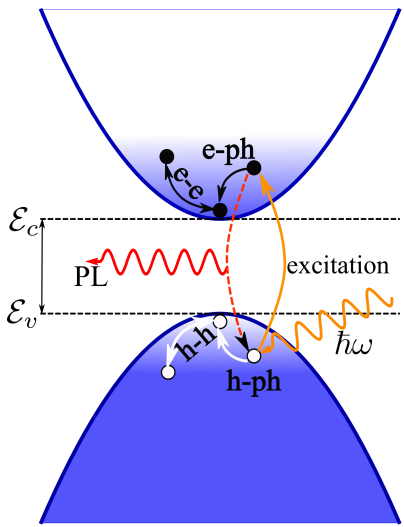}
\caption{Schematic picture of the band structure of a representative semiconductor (GaAs) and the different processes participating in determining the non-equilibrium steady-state distributions. $\mathcal{E}_c$ and $\mathcal{E}_v$ are the energies of the conduction and valence band edges, respectively, and the chemical potential is placed at the middle of the band-gap. Orange and red arrows denote photo-excitation and recombination (that leads to PL), respectively. $\text{e-e}$ ($\text{h-h}$) solid black (white) arrow denotes carrier-carrier scattering and $\text{e-ph}$ ($\text{h-ph}$) black (white) arrow denotes the carrier-phonon scattering.}
\label{fig_schematic}
\end{figure}

\section{Microscopic Formulation} \label{formulation}
To calculate the NESS carrier distributions $f_e(\E)$ (for electrons) and $f_h(\E)$ (for holes), while keeping the phonon subsystem in its own thermal equilibrium~\cite{Dubi2019}, we use the semi-quantum Boltzmann equations (BEs), 
\begin{eqnarray}
\frac{\partial f_c(\E;T_{ph})}{\partial t} &=& \left( \frac{\partial f_c}{\partial t} \right)_{exc} + \left( \frac{\partial f_c}{\partial t} \right)_{rec} \nonumber \\ &+& \left( \frac{\partial f_c}{\partial t} \right)_{c-ph} + \left(\frac{\partial f_c}{\partial t} \right)_{c-c}, \label{BE-elec} 
\end{eqnarray}
where $f_c$ be the electron distribution for $c = e$ in the conduction band, and hole distribution for $c = h$ in the valence band at an energy $\E$; $T_{ph}$ is the phonon temperature. 

The right-hand-side of the BEs includes four central processes. The first is the photo-excitation process, which occurs upon absorption of a photon of energy $\hbar \omega$. 
The next process is the recombination of excited carriers by which photo-excited electrons and holes lose energy by emitting photons, a process known as PL. Third, the carrier-phonon ($c-ph$) interaction is responsible for transferring energy from the electrons and holes to the lattice. Last but not least, carrier-carrier scattering is responsible for thermalization of the electrons and holes in their respective bands. Going forward, we explain and formulate these four processes in Sections~\ref{sec:ph-exc}-\ref{sec:carrier-carrier}. We neglect re-absorption of spontaneously emitted photons, Auger recombination and carrier-carrier Umklapp scattering, the reasons being explained in Sections~\ref{sec:recombination}, and~\ref{sec:carrier-carrier}, respectively.


\subsection{Photo-excitation}\label{sec:ph-exc}
The first term in Eq.~\eqref{BE-elec}, $\left(\frac{\partial f_c}{\partial t} \right)_{exc}$, denotes the change in the population of carriers in the conduction and valence bands due to absorption of a photon of energy $\hbar \omega$, see Fig.~\ref{fig_schematic}. In semiconductors, following the absorption of a photon with energy $\hbar \omega > \E_g$, the band-gap, an electron-hole pair is created either across the band-gap or inside a specific band. We adapt the formalism introduced in \onlinecite{Kornbluth_NitzanJCHemPhys} to incorporates the band-gap corresponding to semiconductors. The formalism of \onlinecite{Kornbluth_NitzanJCHemPhys} is simple enough, and correctly incorporate the quantum-like term ($\sim |E|^2$, incoherent) in the photo-excitation term of the BE, as has been argued in a previous study in connection to the metal nanoparticles \cite{Dubi2019}. It further bypasses the need for $a~priori$ exact determination of the electron-photon interaction matrix element by using physical conditions, such as the normalization of the total probability of all the electronic processes and the particle number conservation, with an input of the absorption lineshape~\cite{Kornbluth_NitzanJCHemPhys}. The change in carrier distribution at energy $\E$, due to photon absorption is given by
\begin{equation}\label{Eq-exc-c}
\left( \frac{\partial f_c(\E)}{\partial t}  \right)_{exc} = \frac{N_{exc} \phi_{exc}^{(c)}(\E) }{\rho_c (\E)},
\end{equation}
where $\phi_{exc}^{(c)}(\E)$ is the net change in the carrier population at $\E$ in the conduction band (for $c =e$) and valence band (for $c = v$). See Appendices \ref{SI_sec_ph-ext-elec} and \ref{SI-sec-ph-exc-hole} for detailed derivation of the explicit formulae.
The rate of number of electrons excited due to the absorption of photon of energy $\hbar \omega$ is then given by $\frac{dn_c}{dt} = \int \rho_c (\E) \left( \frac{\partial f_e(\E)}{\partial t} \right)_{exc}$, which is equal to $N_{exc}$ electrons per unit time per unit volume. A positive value of $N_{exc}$ signifies an increase in the particle number, i.e., an increase in the number of 
electrons in the conduction band and holes in the valence band. When the semi-conductor is continuously illuminated by light, an ever increasing particle number indicates that the system shall never reach a steady-state until all the electrons (holes) are moved to the conduction (valence) band, thereby losing the semiconducting property. A balancing process that comes to rescue this catastrophe is the spontaneous recombination which we formulate in Sec. \ref{sec:recombination}.


In our formalism of the photo-excitation we do not explicitly incorporate momentum conservation, 
because in the presence of interactions of electrons and holes with phonons, defects and impurities present in the system, as well as for systems with finite size, the momentum is no longer a good quantum number. In this regard, for direct band-gap SCs, our formulation of the photo-excitation process provides an upper bound for the number of excitation events (i.e., the maximum number of photo-excited carriers). 

For indirect band-gap SCs, the interband transition due to photon absorption may require intraband carrier-phonon interaction to provide the necessary momentum transfer. This is because the conduction band minimum and the valence band maximum do not appear at the same $\mathbf{k}$-point in the Brillouin zone while an optical transition is associated with near-zero momentum transfer. Moreover, any inter-valley (intraband optical) transition (for example, 
from the $\Gamma$-valley to $L(\text{or~} X)$-valley) requires a direct intraband optical transition accompanied by a carrier-phonon and/or a carrier-carrier Umklapp scattering. Therefore, due to the absence of explicit momentum conservation our formulation of photo-excitation process can incorporate both indirect interband transitions and inter-valley (intraband) optical transitions only in the sense that it provides the upper bound for the number of excitation events. Nevertheless, for the intensities considered in this study, we explicitly show in Appendix \ref{SI_sec_ph-ext-elec} (equations \eqref{Appdx:small_intrababd} and \eqref{Appx:rhocc_by_rhovc}) that intraband optical transitions are negligible compared to their interband counterpart. We show this to be true irrespective of whether the transition is direct or indirect. However, for even higher intensities and $\hbar \omega \sim \E_g$ intra-band optical transitions should be considered explicitly.  Moreover, how much the explicit incorporation of inter-valley scattering would change the NESS quantitatively at the steady state in our CW illumination case is beyond the scope of present study.



\subsection{Recombination}\label{sec:recombination}
The second term in Eq.~\eqref{BE-elec}, $\left(\frac{\partial f_c}{\partial t} \right)_{rec}$, denotes the interband recombination of excited carriers, by which excited electrons and holes lose energy by spontaneously emitting photons leading to PL, see Fig. \ref{fig_schematic}. We adopt the standard formulation, see Appendices \ref{SI-2-sec:rece} and \ref{SI-2-sec:rech}, for PL,~\cite{photolum1,photoluminescence2,Kamide_PhysRevApplied} which is also consistent with the formulation of the photo-excitation (Section~\ref{sec:ph-exc}) considered above. The rate of change of electron population due to the recombination of carriers with energy $\E$ is given by
\begin{equation}\label{Eq:rec-particle}
\left(\frac{\partial f_c}{\partial t} \right)_{rec} = - \frac{N_{rec} \phi_{rec}^{(c)}(\E)}{\rho_c(\E)},  
\end{equation}
and the rate of change of particles due to recombination is given by $\frac{dn_{rec}}{dt} = \int d\E \rho_c (\E) \left( \frac{\partial f_e}{\partial t} \right)_{rec}$ $ = - N_{rec}$, a negative sign signifying a loss of particles. Here, $\phi_{rec}^{(c)}(\E)$ is the  net change in the carrier population at energy $\E$ due to the recombination. See Appendix \ref{Appdx:Recombination} for details of the formulation. At the steady-state, the rate of photo-excitation of electrons would become the same as that of recombination of electrons, and this fixes $N_{rec} = \frac{dn_e}{dt}$ so that $\frac{dn_e}{dt} + \frac{dn_{rec}}{dt} = 0$. It is worthwhile to point out the recombination rate, therefore, indirectly depends on the photo-excitation rate via $\frac{dn_e}{dt}$.

To this end, we relax the requirement of the momentum conservation for the same set of reasons corresponding to the photo-excitation process. Therefore, our formulation of recombination provides  only an upper limit of the number of recombination events for indirect band-gap SCs.

Furthermore, we assume absorption of spontaneously emitted photons is a weak effect. Such an assumption is valid for systems with size smaller than the optical skin depth, such as nanoparticles and thin films. Most semiconductor applications, such as solar cells, lighting applications~\cite{Kamide_PhysRevApplied,Green2017,Konig20102862,CONIBEER2009713,LED-book,Li2006Semiconductor-Physical-Electronics} etc. meet this assumption. At this system size, the effective local electric field corresponding to the illumination is homogeneous throughout the system, and almost all the spontaneously emitted photons from inside the bulk of the system are radiated out of the system with a negligible fraction getting reabsorbed thus, justifying our assumption. 


\subsection{Carrier-phonon interaction}\label{sec:carrier-phonon}
The electron-phonon ($e-ph$) and hole-phonon ($h-ph$) interactions (third term in Eq.~\eqref{BE-elec}) are responsible for transferring energy from the electrons and holes to the lattice. These are therefore, responsible for the heating of the lattice and cooling of the carriers and shown schematically shown in Fig. \ref{fig_schematic}. They occur within an energy window comparable to the Debye energy near the band edge (thus, it is typically narrow with respect to the photon energy). For simplicity, we consider the deformation potential approximation for the interaction between the carriers (both electrons and holes) and the phonons~\cite{SnokePRB1992}, where only the longitudinal acoustic phonons are taken into account; nevertheless, our formulation can be extended to incorporate optical phonons, both polar and longitudinal~\cite{ridley1999quantum}. In general, a more accurate quantitative estimation of the energy transfer from carriers to phonon in semiconductors would require explicit inclusion of other phonon modes, such as, transverse acoustic, polar optical phonons~\cite{ridley1999quantum} (arguably, known to be the dominant one in III-V semiconductors~\cite{SjakstePhysRevB2018, Bernardi5291}), inter-valley carrier-phonon scatterings~\cite{ridley1999quantum}. However, such a detailed analysis is not the primary motivation of the present study. Our calculations can be considered as an order of magnitude estimation of the carrier-phonon energy transfer and in Sec.~\ref{sec:partition} we show that our formalism agrees well with the experimental predictions in the appropriate regime.

The energy of an electron in the conduction band is measured from the conduction band edge, and $\E = \E_c + \frac{\hbar^2 k^2}{2 m_e^*}$ where $m_e^*$ is the conduction band effective mass of the electron and $\E_c$ is the energy of the conduction band edge. Following Refs.~\cite{ridley1999quantum,SnokePRB1992}, the Fermi golden rule expression for the scattering between the conduction band electrons and phonons is obtained to be
\begin{widetext}
\begin{eqnarray}\label{SI-2-Eq:elec-ph-rate}
\left(\frac{\partial f_e(\E)} {\partial t} \right)_{e-ph} &=& \frac{\Xi_c^2}{4 \pi \rho_c} \frac{\sqrt{m_e^*}}{\sqrt{2(\E - \E_c)}} \int_0^{\hbar \omega_D} \frac{d\E_q~\E_q^2}{(\hbar v_{ph})^4} \Bigg [(n_B(\E_q, T_{ph}) + 1) \Big( f_e(\E + \E_q) [1 - f_e(\E)] \nonumber \\ &-& f_e(\E) [1 - f_e(\E - \E_q)] \Big) + n_B(\E_q, T_{ph}) \Big(f_e(\E - \E_q) [1 - f_e(\E)] \nonumber \\ &-& f_e(\E) [1 - f_e(\E + \E_q)] \Big) \Bigg],
\end{eqnarray}
 \end{widetext} 

where we have assumed that a phonons are in thermal equilibrium characterized by a Bose-Einstein distribution $n_B(\E_q, T_{ph})$ at the lattice temperature $T_{ph}$, with $n_B(\E_q, T_{ph}) = (e^{\E_q/k_B T_{ph}} - 1)^{-1}$, $k_B$ being the Boltzmann constant expressed in eV and $\hbar \omega_D$ is the `Debye' energy. Within the deformation potential approximation the acoustic phonons exhibit a linear dispersion, viz., $\omega_{\q} = v_{ph} |\q|$, $v_{ph}$ being the speed of sound (see Table~\ref{Tab:tableofparam} for the numerical values used)~\cite{Mahan2000}. In~\eqref{SI-2-Eq:elec-ph-rate}, $\Xi_c$ is the deformation potential constant, $\rho_c$ is the material density, $V$ is the volume of the system~\cite{Mahan2000}.

Analogously, the energy of a hole in the valence band is measured from the valence band edge, and $\E = \E_v - \frac{\hbar ^2 k^2}{2 m_h^*}$, where $m_h^*$ is the valence band effective mass of the hole and $\E_v$ is the energy of the valence band edge. Therefore, energy of a hole with momentum $\mathbf{k}$ is $\E_h = \E_v - \E$, i.e., the lower the value of $\E$ the more energy the hole exhibits. A Fermi golden rule expression corresponding to the hole-phonon interaction, analogous to the electron-phonon one, is given by,
\begin{widetext}
\begin{eqnarray}\label{SI-2-Eq:hole-ph-rate}
\left(\frac{\partial f_h(\E)} {\partial t} \right)_{h-ph} &=& 
\frac{\Xi_v^2}{4 \pi \rho_m} \frac{\sqrt{m_h^*}}{\sqrt{2(\E_v-\E)}} \int_0^{\hbar \omega_D}  \frac{d\E_q~\E_q^2}{(\hbar v_{ph})^{4}} \Bigg [(n_B(\E_q, T_{ph}) + 1) \Big( f_h(\E - \E_q) [1 - f_h(\E)] \nonumber \\ &-&  f_h(\E) [1 - f_h(\E + \E_q)] \Big) + n_B(\E_q, T_{ph})\Big( f_h(\E + \E_q) [1 - f_h(\E)] \nonumber \\ &-&  f_h(\E) [1 - f_e(\E-\E_q)] \Big) \Bigg],
\end{eqnarray}
\end{widetext} 
where $\Xi_v$ is the deformation potential for the $h-ph$ interaction corresponding to the holes in the valence band (see Table \ref{Tab:tableofparam}).

\subsection{Carrier-carrier interaction}\label{sec:carrier-carrier}
The electron-electron interaction in the conduction band and hole-hole interaction in the valence band (fourth term in Eq.~\eqref{BE-elec}, altogether carrier-carrier scattering) are responsible for thermalization of the electrons and holes in their respective bands. This enables carriers to reach a quasi-equilibrium where a Fermi-Dirac distribution with a carrier temperature [$T_{e(h)}$ for electrons (holes)] can be associated to the carriers. 

Following the standard notations from the \onlinecite{SnokePRB1992}, we obtain the net scattering rate corresponding to the electrons with energy within $\E$ and $\E + d\E$ is given by
\begin{widetext}
\begin{align}
&\left(\frac{\partial f_e(\E)}{\partial t}\right)_{e-e} = \frac{4\pi}{\hbar} \frac{V^2}{16\pi^{4}}\int d\E_1d\E_2d\E_3 \left[\int_{k_{low}}^{k_{high}} d\kappa |M(\kappa)|^2 \right] \delta(\E_1 -\E_2 +\E -\E_3) \times \nonumber \\ & \frac{\hbar}{\sqrt{(\E - \E_c)}} \left( [1 - f_e(\E)][1 - f_e(\E_1)]f_e(\E_2)f_e(\E_3) - f_e(\E)f_e(\E_1)[1 - f_e(\E_2)][1-f_e(\E_3)] \right), \label{SI-2-Eq:elec-elec-rate}
\end{align}
\end{widetext}
where $V$ is the volume of the system under consideration and $M(\kappa)$ is the matrix element of the electron-electron interaction Hamiltonian within the first-order perturbation theory; it depends on the momentum exchange, $\kappa = \mathbf{k} - \mathbf{k}_2 = \mathbf{k}_1 - \mathbf{k}_3$. The limits of the integration over the matrix element are $k_{low} = \text{max}(|k-k_2|,|k_1 - k_3|)$ and $k_{high} = \text{min}(k+k_2 , k_1 + k_3)$ where the electron wave-number $k_j$ corresponds to $\E_j$.

We consider the short-ranged, hard sphere interaction between the electrons for which $\int_{k_{low}}^{k_{high}} d\kappa |M(\kappa)|^2 = \frac{\sigma_t (m_e^*)^2}{4 \pi^2 \hbar^3 V^2} (k_{\text{high}} - k_{\text{low}})$, where $\sigma_t = 6\times 10^{-16}~ m^2$ is the total scattering cross-section corresponding to electrons in GaAs~\cite{SnokePRB1992}. 
Notice the fact that the matrix element $M^2(\kappa)$ contains a $V^{-2}$ factor which cancels the $V^2$ appearing in~Eq. \eqref{SI-2-Eq:elec-elec-rate}, thereby making the expression for electron-electron scattering volume normalized. We point out in passing that our choice of the hard-sphere scattering is a trade-off between the computation time and a more realistic model of screened Coulomb interaction, the Debye screening \cite{Ashcroft-Mermin, ridley1999quantum}. In \onlinecite{SnokePRB1992, Snoke_PhysRevB,Snoke_PhysRevB.47.13346} it has been pointed out that at low densities ( $n_{e}< 10^{22} m^{-3}$) the ``Debye" screening formula \cite{Ashcroft-Mermin, ridley1999quantum} breaks down and electrons act as unscreened particles. Therefore, following \onlinecite{SnokePRB1992} we take hard-sphere interaction with the total scattering cross-section $\sigma_t $ representing the value of the scattering cross-section of unscreened electrons. 

An expression for the hole-hole interaction in the valence band, analogous to the electron-electron interaction term, is given by
\begin{widetext}

\begin{align}
&\left(\frac{\partial f_h(\E)}{\partial t}\right)_{h-h} = \frac{4\pi}{\hbar} \frac{V^2}{16\pi^{4}}
\int d\E_1d\E_2d\E_3 \left[\int_{k_{low}}^{k_{high}} d\kappa|M(\kappa)|^2 \right] \delta(\E_1 -\E_2 +\E -\E_3) \times \nonumber \\ & \frac{\hbar}{\sqrt{(\E_v - \E)}} \Big( [1 - f_h(\E)][1 - f_h(\E_1)]f_h(\E_2)f_h(\E_3) - f_h(\E)f_h(\E_1)[1 - f_h(\E_2)][1 - f_h(\E_3)] \Big).\label{SI-2-Eq:hole-hole-rate}
\end{align}
\end{widetext}
 
The hard sphere interaction between the holes is given by $\int_{k_{low}}^{k_{high}} d\kappa |M(\kappa)|^2 = \frac{\sigma_t (m_h^*)^2}{4 \pi^2 \hbar^3 V^2} (k_{high} - k_{low})$, where $\sigma_t = 6\times 10^{-16}~ m^2$, being equal to that of the electrons, is the total scattering cross-section corresponding to holes in GaAs. 

\textcolor{black}{The quantum nature of the carrier-carrier scattering comes from the Pauli exclusion principle incorporated into the distribution-dependent term.
The carrier-carrier scattering rate depends on the density of (excited) carriers, and therefore, on the intensity of the incident light and the carrier temperature~\cite{Snoke_PhysRevB,Kash_PhysRevB}. For carrier densities $n_e <10^{24}~m^{-3}$, we replace the the screened Coulomb interaction by the hard sphere interaction involving a total carrier-carrier scattering cross-section that determines the strength of carrier-carrier scattering ~\cite{Snoke_PhysRevB,Snoke_PhysRevB.47.13346,Snoke2011_ann_physik}. In this carrier density regime screening energy, $\E_{TF} = \frac{\hbar^2 \kappa_{TF}^{2}}{2m_{e}^{*}}$ with $\kappa_{TF}$ being the Thomas-Fermi wave-vector \cite{Ashcroft-Mermin}, always remain much smaller than the average energy of the electrons which validates the use of hard-sphere interaction (this applies to the holes too). However, in this situation carrier-carrier exchange interaction also become important \cite{snoke_2020, SNOKE20121825} but within hard-sphere interaction it can be incorporated within an adjustable parameter in the total carrier-carrier scattering cross-section \cite{snoke_pvt}. In most of the undoped III-V binary SCs with wider band-gaps [with exceptions being InSb ($\E_g = 0.235~ eV$), InAs ($\E_g = 0.417~ eV$) at room temperature] carrier density does not reach such high values for the illumination intensities considered in our study~\cite{Becker_PhysRevLett_1988,Snoke_PhysRevB,Snoke_PhysRevB.47.13346}. Therefore, our formulation of carrier-carrier interaction applies to a large class of III-V binary SCs. For Ternary and Quaternary alloys energy band-gaps and room temperature carrier density strongly depend on the alloy composition~\cite{Vurgaftman_param_2001}, and therefore, these are needed to be studied case by case.} Importantly, we discard Auger (non-radiative) recombination here, as it is negligible for the intensities we consider~\cite{Strauss_APL_1993_Auger,Johnson_JVST_2007_Auger,GovoniPhysRevB.84.075215}, and become appreciable only under extremely high carrier densities with spatial confinement (such as small nano-structures with system size even much less than that we already assumed in connection to absorption of the spontaneously emitted photons in Section~\ref{sec:recombination}) where spatial confinement can lead to a large overlap between electron and hole wave functions~\cite{Achermann2006}. Moreover, we consider only pristine SCs, such that carrier-impurity scattering is neglected~\cite{SzeCh1}.


\section{Macroscopic formulation - Energy and number conservation}\label{sec:macro}
Crucially, in order to account correctly for the non-equilibrium properties, energy conservation must be considered for each sub-system (electrons, holes and phonons) separately~\cite{Dubi2019}.
The rate of change of energy of the electronic subsystem (and analogously for holes) is obtained by integrating over the distribution functions~\eqref{BE-elec} 
by $\int d\E (\E - \E_C) \left(\frac{\partial f_e}{\partial t} \right) \rho_e(\E)$, $\E_C$ being the energy of the conduction band edge $\Big(\int d\E (\E_V - \E) \left(\frac{\partial f_h}{\partial t} \right) \rho_h(\E)$, $\E_V$ being the energy of the valence band edge$\Big)$ [see Fig. \ref{fig_schematic}] 
, resulting in 
\begin{equation}\label{Ue}
\frac{d \mathcal{U}_c}{dt} = W_{c-exc} - W_{c-rec} - W_{c-ph},
\end{equation}
where $\rho_c(\E)$ is the carrier density of states (cDOS). 
We interpret equation~\eqref{Ue} as the balance between the rate of gain of energy of the electronic (hole) subsystem due to the photo-excitation, viz., $W_{c-exc}$, the rate of loss of excess energy the electrons (holes) due to they recombination back to the conduction (valence) band, viz., $W_{c-rec}$, and the rate of energy flow to the lattice, $W_{c-ph}$. The elastic carrier-carrier scattering processes do not cause any change in the energy of the carrier subsystems, respectively. At the steady-state $\frac{d \mathcal{U}_c}{dt} = 0$ signifies the conservation of energy. See Appendix \ref{SI-sec:energy-conserve} for the detailed expressions. 

Similarly, the total energy of the lattice, $\mathcal{U}_{ph}$, is balanced by the heat flowing in from the electron and hole subsystems and flowing out to the environment, viz.,
\begin{equation}\label{Uph}
\frac{d \mathcal{U}_{ph}}{dt} = (W_{e-ph} + W_{h-ph}) - G_{ph-env} (T_{ph} - T_{\text{amb}}),
\end{equation}
where $G_{ph-env}$ is the coupling between the lattice and the environment, and is phenomenologically introduced in the formulation.See Appendix \ref{SI-sec:energy-conserve} for the detailed expressions.
Moreover, value of $G_{ph-env}$ may strongly depend on the geometry of the sample and on the thermal conductivity of the host~\cite{Dubi2019,Dubi-Sivan-Faraday}.

We search numerically for a NESS based on the energy conservation, i.e., $\frac{d \mathcal{U}_e}{dt} = \frac{d \mathcal{U}_h}{dt} = \frac{d \mathcal{U}_{ph}}{dt} = 0$, when the semiconductor is maintained under CW illumination. When a steady-state is obtained, equations~\eqref{BE-elec}-\eqref{Uph} provide us with the electron and hole distributions corresponding to the NESS,
and the temperature of the lattice, $T_{ph}$. We then extract the steady-state electron and hole temperatures, $T_e$ and $T_h$, respectively, by adapting two generic experimental procedures frequently used, (i) from the exchange of power between the electron (hole) sub-systems and the phonons~\cite{dubi2011colloquium,cui2017study}, and (ii) from the steady-state PL spectra~\cite{HirstInGaAs-PL,Nguyen2018_Nature_Energy,Giteau-JAP2020}.

The physical picture of the power transfers corresponding to the energy conservation is the following. The incident radiation excites electrons in the conduction band and holes in the valence band, thereby generating high energy (non-thermal) carriers in the respective bands. To reach a steady-state, electrons in the conduction band then lose their energy chiefly by (i) emitting phonons (via electron-phonon ($e-ph$) interactions), and (ii) by recombining with the holes in the valence band. Holes too follow the same processes (but recombine with electrons in the conduction band). Under continuous illumination, given $e (h)-ph$ interactions are intra-band processes, these alone cannot force the system to reach a steady-state, so that recombination turns out to be important as well. There are other processes, such as intra-band Auger recombination, impact ionization etc., however, the rates of these are known to be much slower compared to the $e-ph$ interactions and the interband recombinations~\cite{Sjakste_2018JPCM, BeckerPhysRevLettFemtosecondPhotonEchoes}. The intraband carrier-carrier scattering is even slower a process in the case of low density of electrons (and holes) in the conduction (and valence) band. For GaAs carrier-carrier interaction plays an active role only at carrier densities $n_e = n_h = 10^{24}~ m^{-3}$~\cite{Sjakste_2018JPCM,BeckerPhysRevLettFemtosecondPhotonEchoes}. A steady-state in the electron and hole gases is reached when $\frac{d \mathcal{U}_e}{dt} = 0 $ and $\frac{d \mathcal{U}_h}{dt} = 0 $, respectively. Energy transferred to the lattice then dissipates to the environment maintained at the ambient temperature $\approx 297$ K in our case, and the lattice phonons reach a steady-state when $\frac{d \mathcal{U}_{ph}}{dt} = 0$.

\section{Results}\label{sec:results}
We take intrinsic (undoped) Gallium Arsenide (GaAs) as a specific material to understand and determine the non-equilibrium carrier distributions under CW illumination with varying intensities. However, our formulation can well be applied to pulsed-illumination with a suitable modification in the photo-excitation term $\left( \frac{\partial f_c}{\partial t} \right)_{exc}$. 
The valence band edge, conduction band edge and the chemical potential are at $\E_V = 3.49$ eV, $\E_C = 4.91$ eV and $\mu = 4.2$ eV, respectively. The illumination photons have energy of $\hbar \omega = 1.65$ eV and the GaAs permittivity is set to $\epsilon''(\omega) = 3.3\epsilon_0$ at $\hbar \omega$~\cite{Aspnes_dielectric_imagine}, where $\epsilon_0$ is the free-space permittivity. The intrinsic carrier concentration in the GaAs at room temperature is $n_e^{\text{amb}} = 1.317 \times 10^{12}~/m^3$, which is considered to be quite low compared to many other semiconductors
~\cite{Goodwin2005,Shur_handbook1}. Other parameters used in this study are given in Table \ref{Tab:tableofparam}.

\begin{table*}
\begin{tabular}{ |c|c|c| }  
\hline
parameter & parameter's symbol & value  \\ 
\hline
effective mass of electron & $m_e^*/m_e$ & 0.063 \cite{zee_ch1} \\ 
effective mass of hole & $m_h^*/m_e$ & 0.49 \cite{zee_ch1}  \\ 
band-gap & $\E_g$ & 1.42 eV
\cite{zee_ch1} \\ 
chemical potential & $\mu$ & 4.2 eV\\ 
bottom of valence band & $\E_b$ & $-\mu/10$\\ 
top of conduction band & $\E_t$ & $2\mu + \mu/10$ \\ 
imaginary part of the dielectric function & $\epsilon''(\omega)$ & 3.3$\epsilon_0$~\cite{Aspnes_dielectric_imagine}\\ 
Debye frequency & $\omega_D$ & 0.0243 eV\\ 
deformation potential in conduction band & $\Xi_c$ & 7.04 \cite{Vurgaftman_param_2001} \\ 
deformation potential in valence band & $\Xi_v$ & 3.6 \cite{Reinhard_hph_1995} \\ 
sound velocity  & $v_{ph}$ & 3650 m/s\\ 
ambient temperature  & $T_{amb}$ & 297 K\\ mass density  & $\rho_m$ & 5320 $\text{kg}/\text{m}^3$ \\ 
free space permittivity & $\epsilon_0$ & $8.8542\times 10^{-12} ~ F/m$\\ 
lattice-environment coupling & $G_{ph-env}$ & $5\times 10^{14}~ W/m^{3}K$ \\
\hline
\end{tabular}
\caption{Values of parameters corresponding to intrinsic (undoped) GaAs. The chemical potential $\mu$ is completely arbitrary and is used in fixing the centre of the band-gap only.}\label{Tab:tableofparam}
\end{table*}

We reiterate that we intend to calculate steady-state distributions and as a first step we consider the parabolic band dispersion with the corresponding carrier density of stats [cDOS] given by,
\begin{eqnarray}\label{DOS} \rho_{e} (\E) &=& \frac{(2m_{e}^{*})^{3/2}}{2\pi^2 \hbar^3}  \sqrt{\E -\E_c}, ~~\text{for}~\E >\E_c, \nonumber \\  \rho_{h} (\E) &=& \frac{(2m_{h}^{*})^{3/2}}{2\pi^2 \hbar^3}  \sqrt{\E_v -\E}, ~~\text{for}~\E <\E_v, \end{eqnarray}
where $m_e^{*}$ and $m_h^{*}$ represent the effective band masses for electrons and holes, respectively. In general realistic band structures of almost all the SCs are far from being parabolic, however for suitable illumination frequencies, for example for GaAs $\hbar \omega \leq \E_{g} + 0.3~eV = 1.72 ~eV $ our results are exact.

However, we point out that our formulation is general enough to accommodate any dispersion relation as long as the cDOS is a function of energy. It is indeed the case with cDOS, be it experimentally obtained or calculated from {\sl ab-initio} density functional theory. Moreover, our formulation can well be generalized to a more complicated band dispersion by incorporating a $\mathbf{k}\cdot\mathbf{p}$ perturbation theory~\cite{ridley1999quantum}. 


\subsection{Non-equilibrium distribution:}\label{sec:distribution} 
\begin{figure*}
\centering
\includegraphics[keepaspectratio=true,scale=0.155]{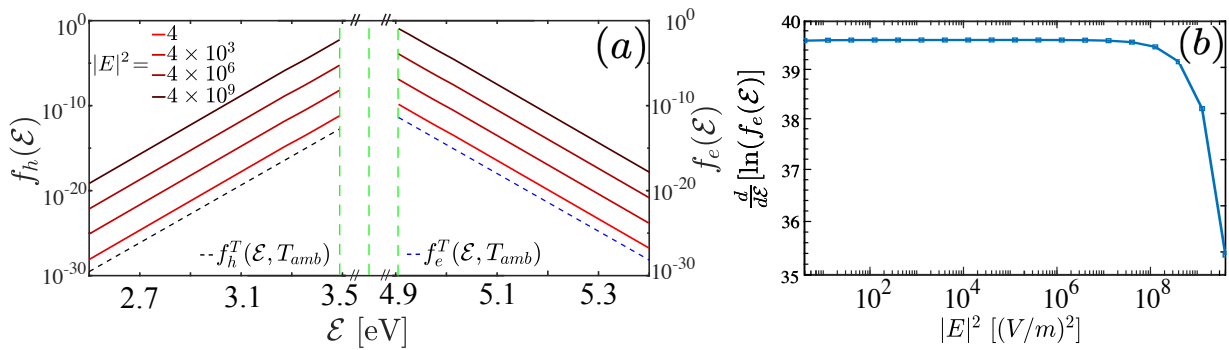}
\caption{\textbf{Hole and the electron distributions:} (a) Plots of the hole distribution $f_h(\E)$ in the valence band, the electron distributions $f_e(\E)$ in the conduction band 
at different values of the local field. 
The valence band edge at $\E_V$, the conduction band edge at $\E_C$ and the chemical potential $\mu$ are shown by green dashed vertical lines. The superscript $T$ denotes the Fermi-Dirac distribution. (b)  Logarithmic slope of the non-equilibrium distribution of electrons $ f_e(\E)$ at electronic energy $\E = 4.914$ eV (an energy near the conduction band edge) as a function of local field $|E|^2$.}
\label{fig_dists}
\end{figure*}

FIG.~\ref{fig_dists}(a) shows the steady-state distributions 
of holes $f_h$ 
and electrons $f_e$ as a function of the energy $\E$ in the valence and conduction bands, respectively, for 
electric field levels ranging from $|E|^2 = 4$(V/m)$^2$ [$1.976\times10^{-6}$ W/cm\textsuperscript{2}] to $|E|^2 = 4 \times 10^9$(V/m)$^2$ [$1.976\times10^{3}$ W/cm\textsuperscript{2}]. For comparison, we plot the thermal (Fermi-Dirac) distributions for electrons and holes $f_e^T$ and $f_h^T$ at ambient temperature. 

The population of the carriers increases with the illumination intensity while the (logarithmic) slopes of the steady-state distribution are nearly the same as that of the ambient thermal distribution at lower intensities.
The common textbook interpretation is that the carrier temperatures do not change at all, and only the effective chemical potential for each of the carriers are shifted to their respective band edges, viz., for electrons towards the conduction band edges and for holes towards the valence band edges \cite{SzeCh1, Ashcroft-Mermin}. However, in the CW illuminated SC the distribution can only be thermalized via carrier-carrier scattering. Due to the negligible carrier-carrier scattering the distributions are indeed not thermalized. Therefore, one can interpret the distributions in the following alternative way.
Since for a thermalized (i.e., Fermi-Dirac) distribution the slope is inversely proportional to the carrier temperature, an increase of the distribution without a change in slope is indicative of a non-thermal distribution. It is only at higher intensities, when $e-e$ and $h-h$ interactions become comparable to the $e(h)-ph$ coupling, that the slopes of $f_{e(h)}$ start to deviate from $f_{e(h)}^T(\E,T_{amb})$. To demonstrate this, in FIG.~\ref{fig_dists}(b) we plot the slopes of $f_e(\E)$ at $\E = 4.914$ eV (near the conduction band edge, where the strength of the $e-e$ interaction is the maximum and therefore, can affect $f_e(\E)$ the most), as a function of the local field $|E|^2$. The slope of $f_e(\E)$ remains unchanged, until the intensity reaches a critical value such that the $e-e$ interaction becomes comparable to the $e-ph$ interaction. This will also be reflected in the carrier temperature (see discussion below and Fig.~\ref{fig_delT}(b)). Thus, at low intensities due to inefficient carrier-carrier interaction, the distributions are truly non-thermal, 
and only start to thermalize at higher illumination intensities. \textcolor{black}{This conclusion is independent of incident photon energies, as demonstrated in Appendix \ref{sec:slope_pht}.}

\subsection{Carrier temperatures} \label{sec:temperatures}
\begin{figure}
\includegraphics[scale=0.395]{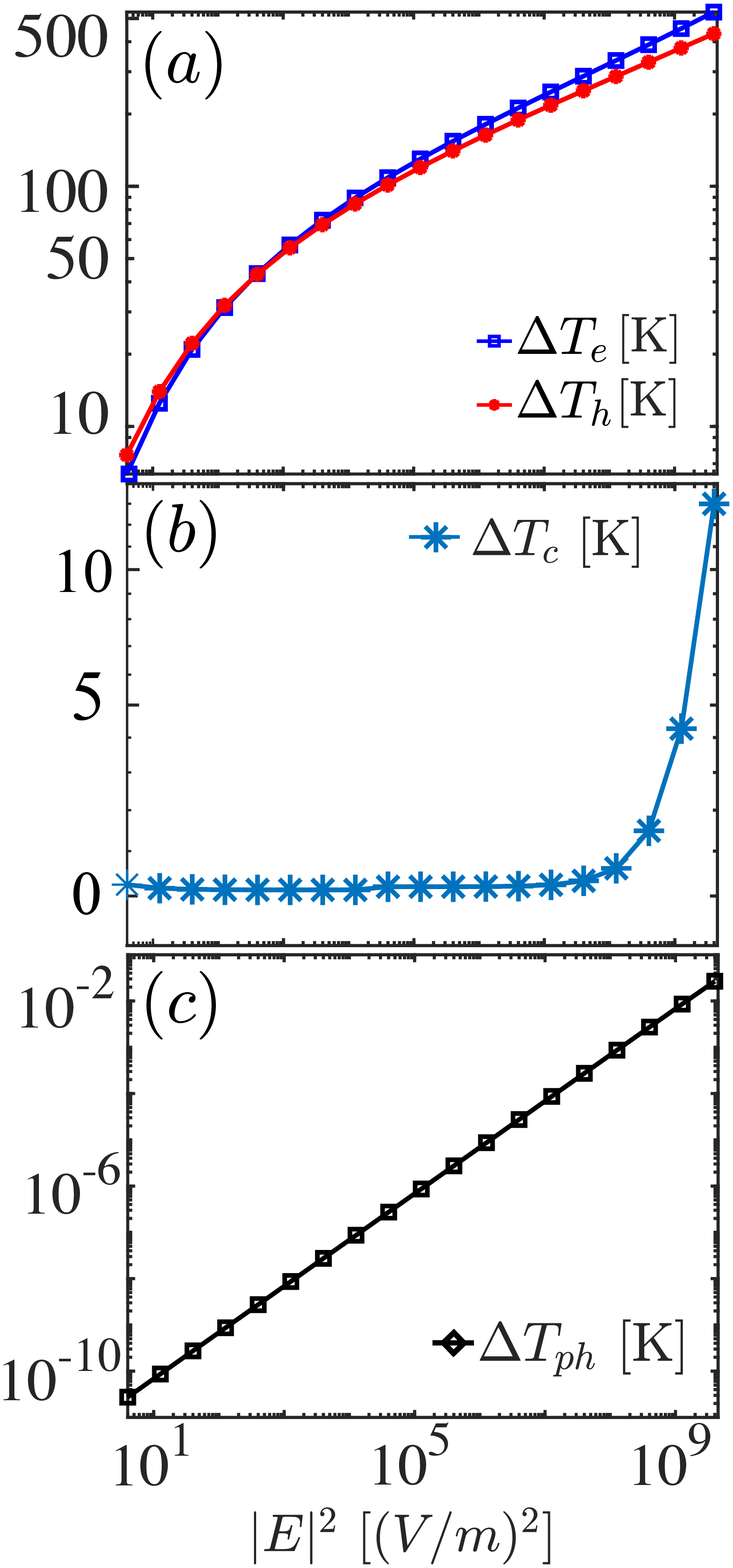}
\caption{\textbf{Temperatures and energy partitions:} (a) The plot of $\Delta T_e = T_e - T_{\text{amb}}$ and $\Delta T_h = T_h - T_{\text{amb}}$ corresponding to the rise in the electron and hole temperatures, respectively, from the ambient temperature as a function of the local field $|E|^2$, in the units of $(V/m)^2$, on a log-log scale obtained from $e-ph$ interaction. (b) The plot of $\Delta T_c = T_c - T_{\text{amb}}$ carrier temperatures from the ambient temperature as a function of $|E|^2$, in the units of $(V/m)^2$, on a log-log scale obtained from fitting the PL spectra, with a $95 \%$ fitting accuracy. (c) The plot of $\Delta T_{ph} = T_{ph} - T_{\text{amb}}$ the rise in phonon temperature from the ambient temperature, indicating the lattice heating, as a function of $|E|^2$ on a log-log scale. The values of $\Delta T_{ph}$ indicates that the lattice does not heat up at all. 
}
\label{fig_delT}
\end{figure}

We now use the NESS distributions to extract macroscopic characteristics of interest. First, we extract (``effective'') electron and hole temperatures from the carrier-phonon power transfers. See Appendix \ref{SI-2-Sec:delte_Weph} for the details of the formalism. We denote $\Delta T_e$ and $\Delta T_h$ as the deviation of electron and hole temperatures, respectively, from the ambient temperature. These are plotted as a function of the local field in FIG.~\ref{fig_delT}(a), showing a nonlinear increase with the illumination intensity by several hundreds of degrees. Such a nonlinear dependence stems from the fact that $W_{e(h)-ph}$ exhibits a nonlinear dependence on $T_{e(h)} - T_{ph}$. The difference between the values of $\Delta T_e$ and $\Delta T_h$ is due to the different effective band masses of electrons and holes, and the different deformation potentials in the conduction and valence bands, see Table \ref{Tab:tableofparam}.

FIG.~\ref{fig_delT}(b) shows the rise in the carrier temperature above ambient, denoted by $\Delta T_c$, extracted from the PL (see Appendix \ref{SI-2-Sec:ss-pl}), as a function of the local field. It is worth emphasizing that the rise in electron and hole temperatures, obtained from the PL spectra is the same, because the recombination process that gives rise to PL is symmetric with respect to electrons and holes. $\Delta T_c$ exhibits a nonlinear increase with the intensity, and differs considerably from $\Delta T_{e(h)}$ both qualitatively and quantitatively throughout the entire range of the local fields.  In particular, $T_c$ obtained from the PL spectra deviate $5-15$~K from $T_{amb}$ which is much smaller compared to the deviation of $T_{e(h)}$ from $T_{amb}$ obtained from carrier-phonon power transfers. However, as explained above, with increasing intensity, thermalization becomes more important, especially above a critical intensity (corresponding to a carrier density $n_e \approx 5.2 \times 10^{19}~ m^{-3}$ here).  
The value of $\Delta T_c$ starts increasing above this critical intensity indicating the role of thermalization in a PL-based temperature measurement. This is also evident from the similarity between FIG.~\ref{fig_delT}(b) and FIG.~\ref{fig_dists}(b). 

In FIG.~\ref{fig_delT}(c) we plot the increase in phonon (lattice) temperature from the ambient, $\Delta T_{ph}$, as a function of the local field, showing a linear dependence on the square of the local electric field. The total power transferred to the lattice from both electron and hole sub-systems exhibits the same linear dependence on the square of the local field (
not shown). At the steady-state we find that $\Delta T_{ph} = (W_{e-ph} + W_{h-ph}) / G_{ph-env}$, explaining the linear dependence of $\Delta T_{ph}$ on $|E|^2$, see Eq.~\eqref{Uph}. The tiny increase in $T_{ph}$ with respect to the ambient temperature implies that the lattice hardly heats up at all. 
However, since $\Delta T_{ph} \propto G_{ph-env}^{-1}$, weaker lattice-environment coupling would lead to more heating of the lattice and vice versa. Interestingly, in the case of metals, electron-phonon coupling $G_{e-ph}$ was orders of magnitude higher due to the larger number of electrons, and even exceeded $G_{ph-env}$ ~\cite{Dubi2019}. 
In that sense, while the bottleneck of the energy flow in metals was the heat transfer from the phonons to the environment, for semiconductors, the bottleneck is the $e(h)-ph$ coupling.

\subsection{Energy partition and efficiency}\label{sec:partition}
Our formulation provides a unique prediction for the partition between the two channels by which the electron/hole sub-system dissipates its energy, viz., interband recombination and energy transfer to the lattice via phonons. This is crucial for the correct prediction of semiconductor heating and associated photo-thermal nonlinearlity in SC nanostructures (for example, recently studied in Silicon nanostructures; notably an indirect band-gap SC)~\cite{SWC_Silicon_nlty,SWC_Anapole_nlty}, 
quantification of the strength of PL, applications relying on these quantities such as thermometry and imaging, 
and most importantly, lighting applications of semiconductors.~\cite{Cognet-phtthm-meth-AnalyChem,Jaque_Vetrone_review,LED-book}.

\begin{figure}
\includegraphics[scale=0.5]{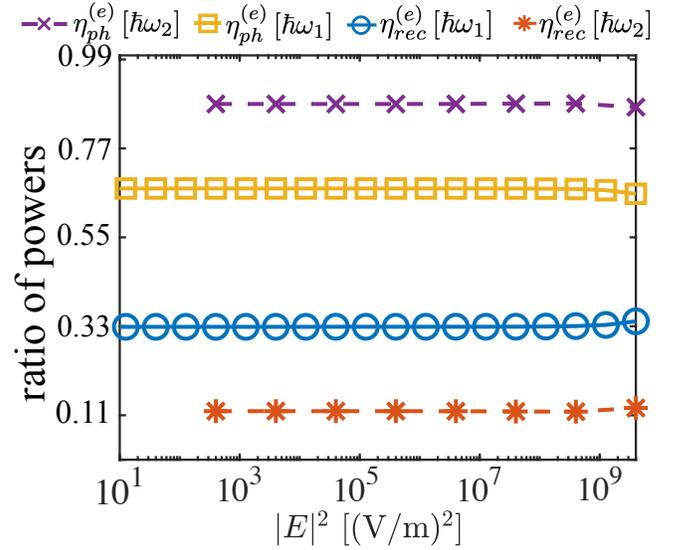}
\caption{\label{eng_part}\textbf{Energy partitions:} The ratios $\eta_{rec}^{(e)}$ (blue circles and red stars) and $\eta_{ph}^{(e)}$ (yellow square and magenta cross) (see text definitions) corresponding to the photon energies $\hbar \omega = 1.65$ eV (yellow and blue solid lines, respectively) and $2.05$ eV (magenta and red dashed lines, respectively), respectively, as a function of $|E|^2$ on a semi-log scale. }
\end{figure}

Thus, we define $\eta_{rec}^{(c)} = W_{c-rec} / W_{c-exc}$ as the ratio of the power dissipated from the electron/hole sub-systems through recombination $W_{c-rec}$ to the power absorbed by the electron/hole sub-systems, $W_{c-exc}$; similarly, $\eta_{ph}^{(c)} = W_{c-ph} / W_{c-exc} [\approx(1 - \eta_{rec}^{(c)})]$ as the ratio between the power transferred from the electron/hole sub-systems to the phonons, $W_{c-ph}$, and the power absorbed by the electron/hole sub-systems. 
FIG.~\ref{eng_part} shows $\eta_{rec}^{(e)}$ and $\eta_{ph}^{(e)}$ for two different photon energies 1.65 and 2.05 eV ($\eta_{rec}^{(h)}$ and $\eta_{ph}^{(h)}$ are quantitatively the same as that of the electrons). We note that the energy partition is inversely proportional to the photon energy, i.e., $\eta_{rec}^{(c)} \propto (\hbar \omega -\E_g)^{-1}$, but does not depend on intensity, see Fig. \ref{eng_part}. \textcolor{black}{To further clarify, we plot $\eta_{rec}^{(e)}$ for four different values photon energy in Fig. \ref{fig_eta_phteng} (a) and (b) in Appendix \ref{Pl_eff_formalism}, which corroborate these conclusions.} The reason is rather straightforward; for higher photon energies, electrons cross the energy gap to higher energies in the conduction band. They, thus, have more energy to lose to phonons before reaching the band edge and recombining. 

Most importantly, the energy partitions can also be used to evaluate the efficiency of the PL process. The total power absorbed $W_{abs}$ by the system is $\hbar \omega N_{exc}$, where $N_{exc}$ is the rate of photon absorption per unit volume, and the total power lost due to PL is $W_{PL} = W_{e-rec} + W_{h-rec} + \E_g N_{exc}$ (note that $N_{exc} = N_{rec}$, the rate of recombination being equal to the rate of excitation at the steady-state, as required by the particle number conservation). From Eq.~\eqref{SI-Eq-We} and Eq.~\eqref{SI-Eq-Wh} we find $W_{e-exc} = \frac{\hbar \omega -\E_g}2 N_{exc} $ and  $W_{h-exc} = \frac{\hbar \omega -\E_g}2 N_{exc}$; these are the powers absorbed by the electron and hole subsystems measured from the conduction and valence band edges, respectively. This signifies that out of the total power absorbed by the semiconductor, viz. $\hbar \omega N_{exc}$, a power of $(\hbar \omega -\E_g)N_{exc}$ is absorbed together by electron and hole subsystems and is equally distributed between them; it also means that a power of $\E_g N_{exc}$ is lost in overcoming the band-gap. $W_{c-rec}$ and $W_{h-rec}$ defined in Eq.~\eqref{SI-Eq-We} and Eq.~\eqref{SI-Eq-Wh}, respectively, measure the power lost by the electrons and holee, respectively. Therefore, the total power lost due to PL is $W_{PL} = W_{e-rec} + W_{h-rec} + \E_g N_{exc}$ which incorporates all possible recombination events from the electrons and holes away from the band edges. Then, the efficiency of PL is defined by $\eta_{PL}^{QE} = \frac{W_{PL}}{\hbar \omega N_{exc}}$, which leads to Eq.~(4). In deriving the final form of Eq.~(4) we define $\eta_{rec}^{(c)} = \frac{W_{c-rec}}{W_{c-exc}}$ [$c=e$ for electrons and $c=h$ for holes] as the ratio of the power dissipated from the electron/hole sub-systems through the recombination $W_{c-rec}$ to the power absorbed by the electron/hole sub-systems, $W_{c-exc}$. 
\begin{eqnarray}\label{PL-eff}
\eta_{PL}^{QE} &=& \frac{W_{PL}}{W_{abs}} = \frac{\eta_{rec}^{(e)} + \eta_{rec}^{(h)}}{2} + \frac{\E_g}{\hbar \omega}\left[1 - \frac{\eta_{rec}^{(e)} + \eta_{rec}^{(h)}}{2} \right]. \nonumber \\
\end{eqnarray}

For the example we study here, we find the efficiency $\eta_{PL}^{QE}$ to be 90.6\% and 73\% at photon energies 1.65 and 2.05 eV, respectively. We compare this to experimental observations~\cite{Johnson_JVST_2007_Auger}, where the PL efficiency was measured as a function of illumination intensity (under laser light of 633nm).
In these experiments, the low-intensity regime is dominated by trap-assisted recombination (e.g., Shockley-Read-Hall recombination), an effect which we do not account for. At high intensities interband recombination is dominant, and the PL efficiency saturates at around 71\% (obtained by interpolating the experimental data), very close to our theoretical result
, showing that the saturation of efficiency comes due to the phonon-mediated dissipation of carrier energy.   

It is worth emphasizing again that although existing formulations for evaluation of PL efficiency account for trap-assisted and Auger recombination~\cite{Hall-SRH-1952,Shockley-Read-SRH-1952,Johnson_JVST_2007_Auger,chuang2009physics,Li2006Semiconductor-Physical-Electronics}, these are macroscopic formulations based on (rate equations of) carrier density instead of carrier distribution and ignore the role of phonons.
This prevents these formulations from being able to explain the high intensity PL efficiency. Put simply, without considering carrier-phonon interactions, PL efficiency should reach 100\% at high intensities, which is in contradiction to experimental observations. 
Our theory thus provides a much improved theoretical prediction for PL efficiency. Moreover, our theory may also serve as a basis for a more rigorous microscopic formulation (in terms of carrier distribution $f_c$) of laser cooling of semiconductors~\cite{PhysRevLett-LaserCooling-2004}, and can allow the quantification of the possible role of carrier-phonon scattering as a heating path-way that hinders cooling~\cite{Morozov2019}.

\section{Conclusions and discussion}  \label{sec: conclusion} 
In conclusion, we employed a semi-quantum BE formulation (including, importantly, energy and particle number conservation), to study the full non-equilibrium carrier distributions and temperatures in an SC under continuous illumination, taking GaAs as a specific example. Our formalism can easily be applied to all the direct band-gap semiconductors, and for indirect semiconductors, it can provide an upper limit of the full non-equilibrium carrier distributions and temperatures. Under low intensity illumination, we find that thermalization processes are inefficient, and the system remains at strong non-equilibrium. Somewhat surprisingly, for high intensities the SC tends to thermalize  more efficiently due to increased carrier-carrier interaction at increased particle number densities. Although the lack of thermalization of the ``hot"  (non-thermal) carriers leaves room for the use of ``hot" carrier- based SC electronics, how much the (more efficient) thermalization (at even higher intensities than considered here) can limit the use of ``hot" carriers remains to be determined. Our theoretical formulation serves as a first step towards such a quantitative estimation.

Our formulation also allowed us to evaluate the steady-state carrier temperatures in two ways, corresponding to two different experimental techniques, namely (i) through carrier-phonon energy transfer (measured via, e.g., a floating thermal probe or a thermocouple~\cite{dubi2011colloquium,cui2017study}), and (ii) photoluminescence-based carrier temperature. We find that these two values deviate substantially from each other, again indicating strong non-equilibrium, and providing direct experimental predictions to test our theory. 

Finally, our formulation provides a simple way to evaluate the the steady-state PL efficiency, which is found to compare remarkably well with experimental observations and goes beyond existing theoretical models~\cite{Hall-SRH-1952,Shockley-Read-SRH-1952,Johnson_JVST_2007_Auger,chuang2009physics,Li2006Semiconductor-Physical-Electronics}. Our model can be used as a theoretical platform to study outstanding open questions in the field, for instance, the persistent long-lived non-thermal carriers and the increase in carrier temperature recently observed in GaAs and other III-V semiconductors under CW illumination~\cite{Tedeschi_acs.nanolett_2016, Shojaei2019, Chen2020, Fast_2020}, and may enable the clarification of the controversial issues related to optical cooling and temperature of SCs~\cite{Morozov2019}. 
Our formalism can be extended to other non-equilibrium situations, such as current-carrying junctions, doped semiconductors, and be used to evaluate and optimize the PL efficiency of, e.g., HC solar cells~\cite{Green2017,Konig20102862,CONIBEER2009713}, optical cooling~\cite{PhysRevLett-LaserCooling-2004,Johnson_JVST_2007_Auger,Roman2020}, and many other light emitting devices~\cite{LED-book,Li2006Semiconductor-Physical-Electronics}. Our formalism can further be fine tuned by incorporating the explicit band structure, inter-valley scatterings, different phonon modes, Auger and impurity scatterings \cite{ridley1999quantum} for a more quantitative estimation of carrier NESS, temperatures and PL efficiency, which we leave for future studies.





\section*{Acknowledgements:} IWU and YS were supported by Israel Science Foundation (ISF) grant (340/2020) and by Lower Saxony - Israel cooperation grant no. 76251-99-7/20 (ZN 3637). Authors acknowledge David W. Snoke for valuable suggestions and discussions.

\onecolumngrid
\appendix
\section{Carrier-photon interaction: photo-excitation}
\subsection{Photo-excitation of electrons} \label{SI_sec_ph-ext-elec}

\begin{figure}
\includegraphics[scale=0.18]{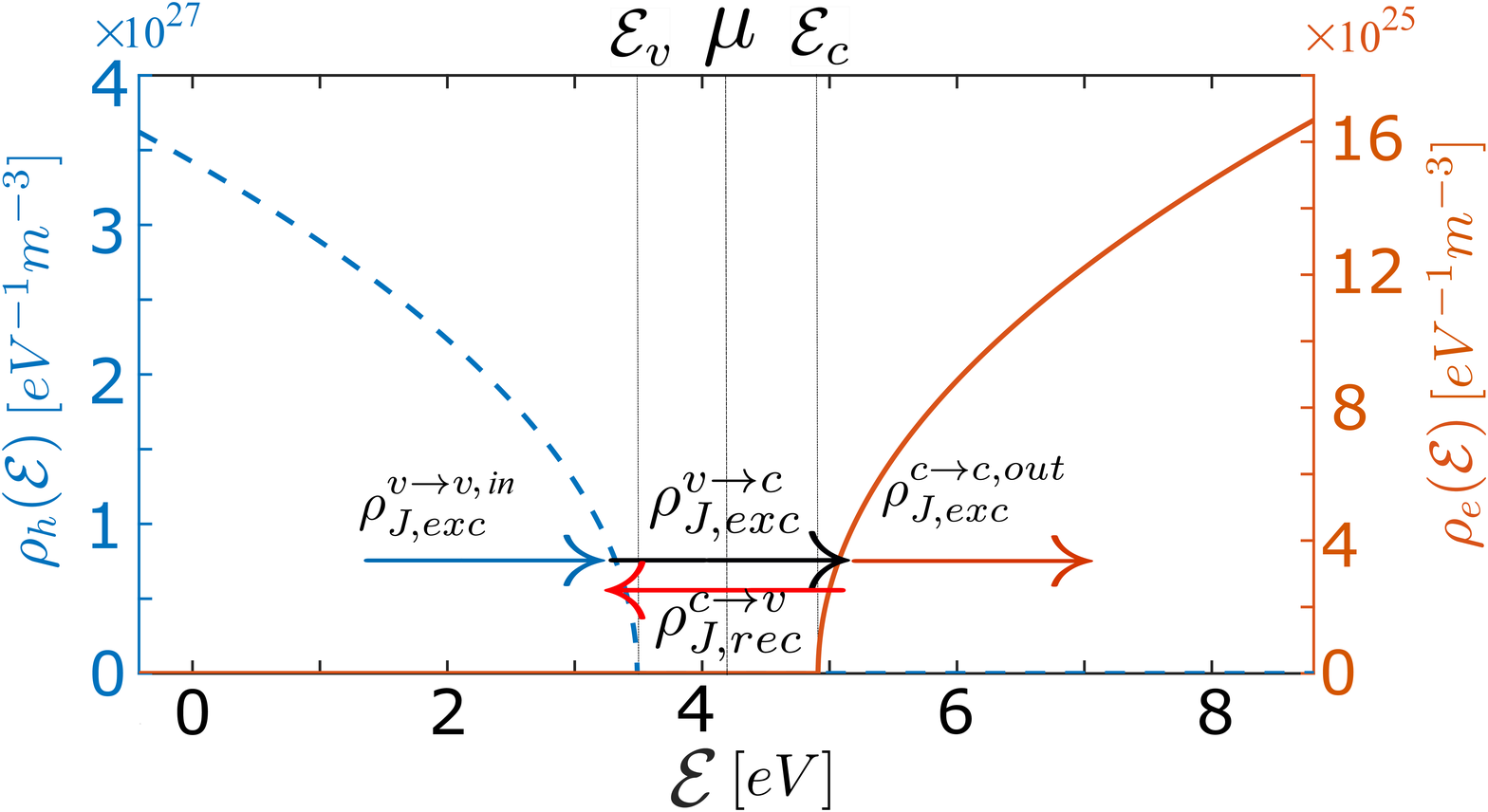}
\caption{Schematic diagram for the excitation and recombination processes. Left $y$-axis plots the hole density of states $\rho_h(\E)$ and right $y$-axis plots the electron density of states $\rho_e(\E)$as a function of energy $\E$, where $\E_c$ denotes the conduction band edge and $\E_v$ denotes the valence band edge. The expression for the joint density of states corresponding to different transitions are explained in the text. Parameters chosen here are for GaAs.
}
\label{fig_exc_rec_schem}
\end{figure}

\paragraph{Interband transition:}
Upon absorption of a photon of energy $\hbar \omega$, an electron-hole pair is created across the band-gap, i.e., if the electron is created in the conduction band at energy $\E$, then the corresponding hole is created in the valence band at energy $\E - \hbar \omega$. The joint probability of absorption of a photon of frequency between $\omega$ and $\omega + d \omega$ and the excitation of an electron in the conduction band at energy $\E$ is given by
\begin{equation}\label{Eq-Avc-elec}
A_{exc}^{v \rightarrow c}(\E_f = \E,\hbar \omega) = K_{exc}^{v \rightarrow c} D_J(\E,\E - \hbar \omega) \rho_{J,exc}^{v \rightarrow c} (\E, \E-\hbar \omega),
\end{equation}
where $D_J(\E_{final},\E_{initial})$ is the square of the matrix element corresponding to the electron-photon interaction leading to the electronic process $\E_{final} \rightarrow \E_{initial}$, $K_{exc}^{v \rightarrow c}$ is the proportionality constant to be defined later, and the superscript `$v \rightarrow c$' denotes the excitation (denoted by the subscript `$exc$') process from the valence band to the conduction band. In Eq.~\eqref{Eq-Avc-elec}, $\rho_{J, exc}^{v\rightarrow c}(\E, \E -\hbar \omega)$ is the joint density of states of the excitation of an electron at $\E$ in the conduction band leaving a hole at $\E - \hbar \omega$ in the valence band, and is given by
\begin{eqnarray}\label{rhovc-elec}
\rho_{J, exc}^{v\rightarrow c}(\E, \E -\hbar \omega) &=& [1 - f_h(\E - \hbar \omega)] \rho_h(\E - \hbar \omega ) \left[1 - f_e(\E)\right] \rho_e(\E), \quad \E > \E_c,
\end{eqnarray}
where $[1 - f_h(\E - \hbar \omega)] \rho_h(\E - \hbar \omega)$ is the electron distribution in the valence band and $[1 - f_e(\E)]\rho_e(\E)$ is the hole distribution in the conduction band, and the density of electron and hole states $\rho_e(\E)$ and $\rho_h(\E)$, respectively, are plotted in Fig.~\ref{fig_exc_rec_schem} (considering the parameters of GaAs). Eq.~\eqref{rhovc-elec} is schematically shown by the black arrow in Fig.~\ref{fig_exc_rec_schem}.

\paragraph{Intraband transition:} The intraband transition, due to the absorption of a photon, can lead to the excitation of an electron-hole pair both within the conduction band and within the valence band with the corresponding probability to be defined below. The joint probability of absorption of a photon of frequency between $\omega$ and $\omega + d \omega$ and the excitation of an electron-hole pair within the conduction band (denoted by superscript $c \rightarrow c$), with the electron at energy $\E + \hbar \omega$ and hole at energy $\E$, is given by
\begin{eqnarray}\label{Eq-Acc-out-elec}
A_{exc}^{c\rightarrow c; out}(\E_f = \E + \hbar \omega, \hbar \omega) = K_{exc}^{c\rightarrow c; out} D_J(\E + \hbar \omega, \E) \rho_{J, exc}^{c\rightarrow c; out}(\E+ \hbar \omega, \E), \quad \E > \E_c,
\end{eqnarray}
where $K_{exc}^{c\rightarrow c; out}$ is the proportionality constant, and the corresponding joint density of state is given by
\begin{equation}\label{rhocc-out-elec}
\rho_{J, exc}^{c\rightarrow c; out}(\E + \hbar \omega , \E ) = f_e (\E) \rho_e (\E ) [1 - f_e (\E + \hbar \omega)] \rho_e (\E + \hbar \omega),
\end{equation} 
and the superscript `$out$' represents the fact that the electron is going out of the state with energy $\E$. The equation~(\ref{rhocc-out-elec}) is schematically shown by the red arrow in Fig.~\ref{fig_exc_rec_schem}. 

Similarly, there exists a finite joint probability for absorption of a photon of frequency between $\omega$ and $\omega + d \omega$, and the excitation of an electron-hole pair within the valence band (denoted by superscript $v \rightarrow v$), with the electron at energy $\E- \hbar \omega$ and hole at energy $\E -2 \hbar \omega$, is given by
\begin{equation}\label{Eq-Avv-in-elec}
A_{exc}^{v \rightarrow v; in}(\E_f = \E - \hbar \omega, \hbar \omega) = K_{exc}^{v \rightarrow v; in} D_J(\E - \hbar \omega, \E- 2\hbar \omega) \rho_{J, exc}^{v \rightarrow v; in}(\E - \hbar \omega, \E - 2 \hbar \omega ), \quad \E > \E_c,
\end{equation}
where $K_{exc}^{v\rightarrow v; in}$ is the proportionality constant, and the corresponding joint density of state is given by
\begin{equation}\label{rhovv-in-elec}
\rho_{J, exc}^{v\rightarrow v; in}(\E - \hbar \omega, \E - 2 \hbar \omega) = [1 - f_h(\E-2 \hbar \omega)] \rho_h(\E - 2\hbar \omega)  f_h (\E - \hbar \omega) \rho_h(\E - \hbar \omega),
\end{equation} 
and the superscript `$in$' represents the fact that the electron is going in to the state with energy $\E - \hbar \omega$ by creating a hole at energy $\E - 2 \hbar \omega$ in the valence band. Equation~\eqref{rhovv-in-elec} is schematically shown by the blue arrow in Fig.~\ref{fig_exc_rec_schem}. 
The above-mentioned joint probability densities corresponding to Eqs.~\eqref{Eq-Avc-elec},~\eqref{Eq-Acc-out-elec}, and~\eqref{Eq-Avv-in-elec} satisfy the following property,
\begin{equation}\label{nomega-elec}
\int_{-\infty}^\infty d\E [A_{exc}^{v \rightarrow c}(\E,\hbar \omega) + A_{exc}^{c\rightarrow c; out}(\E + \hbar \omega, \hbar \omega) + A_{exc}^{v\rightarrow v; in}(\E - \hbar \omega, \hbar \omega)] =\frac{n_{exc}(\omega)}{N_{exc}},
\end{equation}
where $n_{exc}(\omega)$ is the number density of absorbed $\hbar \omega$ photons per unit time between $\omega$ and $\omega + d\omega$, and
\begin{equation}\label{NA-elec}
N_{exc} =  \int d \omega n_{exc}(\omega) = \hbar^{-1} \epsilon^{\prime \prime} (\omega, T_e, T_h, T_{ph}) \langle \mathbf{E}(t) \cdot \mathbf{E}(t)\rangle_t,
\end{equation}
$\langle \cdot \rangle_t$ being the temporal average over a single optical cycle. In the calculation we use $N_{exc} = \frac{2 \epsilon_0 \epsilon''(\omega)}{\hbar}|E|^2$, $\hbar$ being in the units of $(J \cdot s)$, which is the total number density of absorbed photons per unit time (in units of $m^{-3} s^{-1}$). Its value is known from electromagnetic simulations. For simplicity, we consider the effective local electric field corresponding to the illumination to be homogeneous throughout the sample. In that sense, we implicitly assume the system size to be smaller than that of the optical skin-depth of the illumination.

To simplify Eq.~\eqref{nomega-elec} we approximate $D_J(\E_{final},\E_{initial})$ by a constant, and consider all the processes corresponding to Eqs.~\eqref{Eq-Avc-elec},~\eqref{Eq-Acc-out-elec}, and~\eqref{Eq-Avv-in-elec} to be equally probable~\cite{Kornbluth_NitzanJCHemPhys, Dubi2019}
such that $K_{exc}^{v \rightarrow v; in} D_J(\E - \hbar \omega, \E- 2 \hbar \omega) = K_{exc}^{c \rightarrow c; out} D_J(\E + \hbar \omega, \E) = K_{exc}^{v \rightarrow c} D_J(\E, \E- \hbar \omega) = K_{exc}^{(e)}$; the normalization constant (with the superscript `$(e)$' denoting that this normalization corresponds to the excitation of electrons), is determined via the following condition,
\begin{equation}\label{Appdx:small_intrababd}
 K_{exc}^{(e)} \int_{-\infty}^\infty d\E [\rho_{J, exc}^{v\rightarrow c}(\E, \E -  \hbar \omega) + \rho_{J, exc}^{c \rightarrow c; out}(\E + \hbar \omega , \E) 
+ \rho_{J, exc}^{v \rightarrow v; in}(\E - \hbar \omega , \E - 2 \hbar \omega)] = \frac{n_{exc}(\omega)}{N_{exc}}.
\end{equation}
To simplify further, we can consider all the intraband photo-excitations probabilities to be negligible, and we show that $\frac{\rho_{J, exc}^{c\rightarrow c; out}(\E + \hbar \omega , \E)}{\rho_{J, exc}^{v \rightarrow c}(\E, \E -  \hbar \omega)} \ll 1$ and $\frac{\rho_{J, exc}^{v \rightarrow v; in}(\E - \hbar \omega, \E -2  \hbar \omega)}{\rho_{J, exc}^{v\rightarrow c}(\E, \E -  \hbar \omega )} \ll 1$.
\paragraph*{}
Firstly, fro Eqs.~\eqref{rhovc-elec} and~\eqref{rhocc-out-elec}, it is easy to see,
\begin{eqnarray}\label{Appx:rhocc_by_rhovc}
 \frac{\rho_{J, exc}^{c\rightarrow c; out}(\E+ \hbar \omega , \E)}{\rho_{J, exc}^{v\rightarrow c}(\E, \E -  \hbar \omega)} =  \frac{f_e (\E) [1 - f_e (\E + \hbar \omega)] \rho_e (\E + \hbar \omega)}{[1 - f_h(\E - \hbar \omega)] \rho_h (\E - \hbar \omega) [1 - f_e (\E)]}~~ \ll 1,  
\end{eqnarray}
given the fact that for an electron at energy $\E$ in the conduction band, $[1 - f_e (\E)] \approx 1$, $[1 - f_e (\E+\hbar \omega)] \approx 1$, $f_e (\E)\ll1$, and $[1 - f_h (\E - \hbar \omega)] \approx 1$, these are ensured by the presence of the gap because $f_{e(h)} (\E)\ll1$ far away from the chemical potential, a condition that is mostly satisfied for intensities considered in the present study. 
Moreover, the ratio of the density of states in the above equation brings in a factor $(m_e^*/m_h^*)^{3/2} \ll 1~ (0.046$ for GaAs, for example). Similar arguments lead to $\rho_{J, exc}^{v \rightarrow v; in}(\E - \hbar \omega, \E -2  \hbar \omega)/\rho_{J, exc}^{v \rightarrow c}(\E, \E - \hbar \omega) \ll 1$. Therefore, we can rewrite Eq.~\eqref{Appdx:small_intrababd} as, 
\begin{equation}\label{Kexc2}
K_{exc}^{(e)} \int_{-\infty}^\infty d\E \rho_{J, exc}^{v\rightarrow c}(\E, \E - \hbar \omega) = \frac{n_{exc}(\omega)}{N_{exc}}.
\end{equation}
We have explicitly verified these arguments in the numerical calculations for intensities we have considered and found the interband transition to be dominant so that using either~Eqs. \eqref{Appdx:small_intrababd}, or~\eqref{Kexc2} does not change the final result. Finally, the net change in the electronic population at $\E$ in the conduction band is, therefore,
\begin{eqnarray}\label{Eq-phinen-elec}
\phi_{exc}^{(e)}(\E) = \int_0^\infty d \omega  \frac{n_{exc} (\omega)}{N_{exc}} \left[ \frac{\rho_{J, exc}^{v\rightarrow c}(\E, \E -  \hbar \omega )}{\int_{-\infty}^\infty d\E \rho_{J, exc}^{v\rightarrow c}(\E, \E -  \hbar \omega )}\right],~~\E \geq \E_c,
\end{eqnarray}
where the explicit expression for the normalization constant $K_{exc}^{(e)}$, given by Eq. \eqref{Kexc2}, has been used. Therefore, the change in electron distribution at energy $\E$, in the conduction band due to photon absorption is given by
\begin{equation}\label{SI-exc-e}
\left( \frac{\partial f_e(\E)}{\partial t}  \right)_{exc} = \frac{N_{exc} \phi_{exc}^{(e)}(\E) }{\rho_e (\E)},
\end{equation}
which is \eqref{Eq-exc-c} for $c=e$ in the conduction band.


\subsection{Photo-excitation of holes}\label{SI-sec-ph-exc-hole}
We aim to formulate the rate of change of hole population $f_h(\E)$ in the valence band due to photon absorption by adapting the formalism introduced in \onlinecite{Kornbluth_NitzanJCHemPhys}. The population probability of holes in the valence band changes both due to the interband and intra-band transitions. However, following the explanations given in the section~\ref{SI_sec_ph-ext-elec} (see equations~Eqs. \eqref{Appdx:small_intrababd}, \eqref{Appx:rhocc_by_rhovc}, and \eqref{Kexc2}, and nearby discussions), we neglect any intra-band transition due to the absorption of a photon. Therefore, the joint probability of absorption of a photon of frequency between $\omega$ and $\omega + d \omega$ and excitation of a hole in the valence band is given by
\begin{equation}\label{Eq-Acv-hole}
A_{exc}^{c\rightarrow v}(\hbar \omega , \E_f = \E) =   K_{exc}^{(h)} D_J(\E , \E+\hbar \omega ) \rho_{J, exc}^{c\rightarrow v} (\E, \E+\hbar \omega),
\end{equation}
where $K_{exc}^{(h)}$ is the normalization to be defined later and the superscript `$(h)$' denotes that the normalization constant corresponds to the excitation of hole. The joint density states corresponding to Eq. \eqref{Eq-Acv-hole} is
\begin{equation}\label{Eq-rhocv-hole}
 \rho_{J, exc}^{c\rightarrow v}(\E, \E+\hbar \omega) = [1 - f_h(\E)] \rho_h(\E) [1 - f_e (\E+\hbar \omega)]\rho_e (\E + \hbar \omega),
\end{equation}
where a hole at $\E+\hbar \omega$ in the conduction band moves to the valence band at $\E$ upon absorption of a photon of energy $\hbar \omega$, $[1 - f_h (\E)] \rho_h (\E)$ being the electron distribution in the valence band and $ [1 - f_e (\E+\hbar \omega)]\rho_e (\E +\hbar \omega)$ being the hole distribution in the conduction band. Assuming $D_J (\E , \E + \hbar \omega)$ to be a constant, i.e., all hole transitions are equally probable, and absorbing them into the normalization constant $K_{exc}^{(h)}$, we find the joint density of states in Eq. \eqref{rhovc-elec} satisfies the following condition,
\begin{equation}\label{Kexch}
K_{exc}^{(h)} \int_{-\infty}^\infty d\E \rho_{J, exc}^{v\rightarrow c}(\E, \E+\hbar \omega) = \frac{n_{exc} (\omega)}{N_{exc}}. 
\end{equation}
Then the net change in the hole population at $\E$ in the valence band is given by
\begin{eqnarray}\label{Eq-phinhn-hole}
\phi_{exc}^{(h)}(\E) = \int_0^\infty d\omega \frac{n_{exc}(\omega)}{N_{exc}} \left[ \frac{\rho_{J, exc}^{c\rightarrow v}(\E, \E + \hbar \omega)}{\int_{-\infty}^\infty d\E \rho_{J, exc}^{v\rightarrow c}(\E, \E+\hbar \omega)} \right],  ~~\E \leq \E_v.
\end{eqnarray}
Therefore, the rate of change in the hole population in the valence band due to the absorption of a photon is given by
\begin{equation}\label{SI-exc-h}
\left(\frac{\partial f_h (\E)}{\partial t} \right)_{exc} = \frac{N_{exc} \phi_{exc}^{(h)}(\E) }{\rho_h (\E)}
\end{equation}
where $\rho_h(\E)$ the eDOS in the valence band. The number of holes excited due to photon absorption is $\frac{dn_h}{dt} = \int d\E \rho_h(\E) \left( \frac{\partial f_h}{\partial t} \right)_{exc}$ which is equal to $N_{exc}$ holes per unit time. Our formulation thus ensures that the number of photo-excited electrons and holes are the same.

%

\section{Recombination: spontaneous emission}\label{Appdx:Recombination}
\subsection{Recombination of electrons}\label{SI-2-sec:rece}

We adapt the theoretical formalism for the spontaneous emission obtained using the usual Fermi golden rule, as explained in \onlinecite{photolum1}, to make it well-matched with the formulation of the photo-absorption. The rate of change of electron population due to the recombination of electrons with energy $\E$ in the conduction band is given by
\begin{equation}\label{SI-2-Eq:rec-elec}
\left(\frac{\partial f_e}{\partial t} \right)_{rec} = - \frac{N_{rec} \phi_{rec}^{(e)}(\E)}{\rho_e(\E)},  
\end{equation}
and the rate of change of particles due to recombination is given by $\frac{dn_{rec}}{dt} = \int d\E \rho_c (\E) \left( \frac{\partial f_e}{\partial t} \right)_{rec}$ $ = - N_{rec}$, a negative sign signifying a loss of particles. Here, $\phi_{rec}^{(e)}(\E)$ is the  net change in the electronic population at energy $\E$ due to the recombination in the conduction band which is given by
\begin{equation}\label{SI-2-Eq:phirec-elec}
\phi_{rec}^{(e)}(\E) = \int_0^\infty d\omega^{\prime} \frac{n_{rec}^{(e)}(\hbar \omega^{\prime})}{N_{rec}} \left[ \frac{\rho_{J, rec}^{c \rightarrow v} (\E, \E - \hbar \omega^{\prime})}{\int_{-\infty}^\infty d\E \rho_{J, rec}^{c\rightarrow v}(\E -\hbar \omega^{\prime}, \E)}\right],~~ \E \geq\E_c,
\end{equation}
where the joint density of states are given by
\begin{equation}\label{SI-2-Eq:rec_spem_rhoJ}
\rho_{J, rec}^{c\rightarrow v}(\E -\hbar \omega^{\prime}, \E) =  f_e(\E) \rho_e(\E)  f_h(\E - \hbar \omega^{\prime}) \rho_h(\E - \hbar \omega^{\prime}),~~ \E\geq \E_c,
\end{equation}
and is normalized such that \begin{eqnarray}\label{SI-2-Eq:n_rec}
 \int_{-\infty}^\infty d\E K_{rec}^{(e)} \rho_{J, rec}^{c \rightarrow v}(\E -\hbar \omega^{\prime},\E) = \frac{n_{rec}^{(e)} (\omega^{\prime})}{N_{rec}},
\end{eqnarray}
where $n_{rec}^{(e)}(\omega^{\prime})$ is the number density of emitted photons (per unit time, unit volume, and per unit frequency) of energy $\hbar \omega'$ within the range $\omega'$ and $\omega' + d\omega'$. The total number of emitted photons is given by $N_{rec} = \int d\omega^{\prime} n_{rec}^{(e)}(\hbar \omega^{\prime})$ which equals the number of electrons recombining per unit volume. The intraband downward transition induced by photon emission, i.e., the intraband recombination is assumed to be negligible \cite{Kornbluth_NitzanJCHemPhys}. 

We initially choose $n_{rec}^{(e)}(\omega^{\prime})$ by the analytic result of the LHS of Eq.~(\ref{SI-2-Eq:n_rec}) for thermal distributions~\cite{photolum1,photoluminescence2}, namely,
\begin{equation}\label{SI-2-Eq:nrece}
n_{rec}^{(e)} (\omega^{\prime}) = \text{const.}~\rho_{phot}(\hbar \omega^{\prime}) \sqrt{\hbar \omega^{\prime} - \E_g} e^{-\beta_e^*(\hbar \omega^{\prime} - \E_g)}.
\end{equation}
where $\rho_{phot}(\hbar \omega^{\prime})$ is the photonic density of states.
In vacuum $\rho_{phot}(\hbar \omega^{\prime}) = \frac{\sqrt{\epsilon(\omega')}\hbar^2 \omega'^2}{\pi^2 (\hbar c)^3}$ (in units of $(eV \cdot m^3)^{-1}$) where $c$ is the speed of light and $\sqrt{\epsilon(\omega')}$ is the frequency dependent refractive index corresponding to the bulk SC.
Note the $\beta_e^*$ appearing in the above equation represents (the inverse of) an energy scale coming from the energy conservation and aids the convergence in the self-consistent calculation, thereby serving as a parameter for convergence.

It is worthwhile to point out that an alternative choice of $n_{rec}^{(e)}(\omega^{\prime}) = \Theta(\hbar \omega^{\prime} - \epsilon_g) \Theta(2\hbar \omega-\epsilon_g -\hbar \omega^{\prime})$, $\hbar \omega$ being the energy of the incident light, which doesn't use any convergence parameter like $\beta_{e}^{*}$, also brings in the energy conservation.
This indicates that the matrix element $K_{rec}^{(e)}$ is quite insensitive to the choice of $n_{rec}^{(e)}(\omega^{\prime})$ which can be attributed to the condition of the particle number conservation, viz., at the steady-state the number of electrons excited must be equal to the number of electrons recombine. The same holds for the holes too.



\subsection{Recombination of holes}\label{SI-2-sec:rech}
Analogous to the recombination of electrons, the joint probability of recombination of a hole from the valence band with an electron in the conduction band and emission of a photon of energy in the range $\omega^{\prime}$ and $ \omega^{\prime} + d \omega^{\prime}$, similar to that of the photo-excitation, is given by
\begin{eqnarray}\label{SI-2-Eq:rec_spem_hl}
A_{rec}^{v \rightarrow c}(\E_f = \E + \hbar \omega^{\prime} , \hbar \omega^{\prime}) = K_{rec}^{c \rightarrow v} D_J(\E + \hbar \omega^{\prime}, \E) \rho_{J, rec}^{v\rightarrow c}(\E + \hbar \omega^{\prime} , \E),
\end{eqnarray}
with the corresponding joint density of states,
\begin{eqnarray}\label{SI-2-Eq:rec_spem_rhoJ_hl}
\rho_{J, rec}^{v\rightarrow c}(\E + \hbar \omega^{\prime}, \E) = f_h(\E) \rho_h(\E)  f_e(\E + \hbar \omega^{\prime}) \rho_e (\E + \hbar \omega^{\prime}), ~~  \E \leq \E_v, 
\end{eqnarray}
where $K^{c \rightarrow v}_{rec}$ is the normalization. Assuming all the electronic (in this case hole) transitions corresponding to the recombination to be equally probable, i.e., $ D_J(\E, \E+ \hbar \omega^{\prime}) K_{rec}^{v\rightarrow c} =K_{rec}^{(h)}$, (the superscript `$(h)$' denote that this normalization corresponds to the recombination of a hole), Eq. \eqref{SI-2-Eq:rec_spem_hl} satisfy the following condition,
\begin{eqnarray}
    \int_{-\infty}^\infty d\E A_{rec}^{v\rightarrow c}(\E +\hbar \omega^{\prime}, \hbar \omega^{\prime} ) = \int_{-\infty}^\infty d\E K_{rec}^{(h)} \rho_{J, rec}^{v\rightarrow c}(\E + \hbar \omega^{\prime}, \E)= \frac{n_{rec}^{(h)} ( \omega^{\prime})}{N_{rec}},
\end{eqnarray}
which define the normalization constant $K_{rec}^{(h)}$. The total number of emitted photons is given by $N_{rec} =  \int d \omega^{\prime} n_{rec}^{(h)}(\omega^{\prime})$ which is equal to the number of re-combinations happening, where $n_{rec}^{(h)}( \omega^{\prime})$ is the number density of emitted photons of energy $\hbar \omega'$ per unit time per unit frequency of emitted photons. We take the expression for $n_{rec}^{(h)}( \omega^{\prime})$ to be the same as that of the electrons but with $\beta_h^*$ in place of $\beta_e^*$.

Then, the rate of change of electron population due to the recombination of a hole at energy $\E$ in the valence band is given by
\begin{equation}\label{SI-2-Eq:rec-hole}
\left(\frac{\partial f_h (\E)}{\partial t} \right)_{rec} = - \frac{N_{rec} \phi_{rec}^{(h)}(\E)}{\rho_h (\E)},  
\end{equation}
where the net change in the electronic population at energy $\E$ in the valence band due to the recombination of holes with the electrons in the conduction band is given by
\begin{eqnarray}\label{SI-2-Eq:phirec-hole}
\phi_{rec}^{(h)}(\E) = \int_0^\infty d \omega^{\prime}\frac{n_{rec}^{(h)}(\hbar \omega^{\prime})}{N_{rec}}  \left[ \frac{ \rho_{J,rec}^{v\rightarrow c}(\E + \hbar \omega^{\prime} , \E) }{\int_{-\infty}^\infty d\E \rho_{J, rec}^{v\rightarrow c}(\E + \hbar \omega^{\prime}, \E) } \right],~~  \E \leq \E_v.
\end{eqnarray}

\section{Logarithmic slope of the distribution for different photon frequency}\label{sec:slope_pht}

\begin{figure}
\centering
\includegraphics[keepaspectratio=true,scale=0.255]{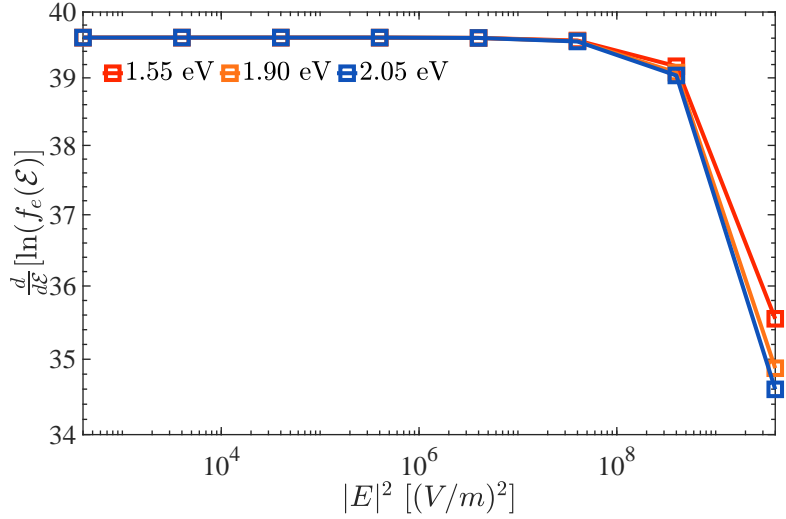}
\caption{Logarithmic slope of the non-equilibrium distribution of electrons $ f_e(\E)$ at electronic energy $\E = 4.914$ eV (an energy near the conduction band edge) as a function of local field $|E|^2$ for different photon energies.}
\label{fig_slope_freq}
\end{figure}

\textcolor{black}{In Sec. \ref{sec:distribution} we demonstrate that due to negligible carrier-carrier scattering the distributions are indeed more non-thermal at low illumination intensities and start thermalizing only at higher intensities due to increased carrier-carrier scattering. Here we demonstrate that this conclusion is independent of the energy of the incident photons, viz., $\hbar \omega$. To demonstrate we plot in Fig. \ref{fig_slope_freq}, the logarithmic slope of $ f_e(\E)$ at electronic energy $\E = 4.914$ eV as a function of local field $|E|^2$ for three more photon energies $\hbar \omega = 1.55$ eV, $1.90$ eV, and $2.05$ eV, respectively. For each $\hbar \omega$ the slope of $f_e(\E)$ remains unchanged, until the intensity reaches a critical value such that the $e-e$ interaction becomes comparable to the $e-ph$ interaction.}

\section{Conservation of energy and number of particles}\label{SI-sec:energy-conserve}
The rates of change of particle numbers, due to the different scattering mechanisms, are given by the following definitions
\begin{eqnarray}
\left(\frac{dn_{e(h)}}{dt}\right)_{exc} &=& \int d\E \left(\frac{\partial f_{e(h)}}{\partial t}\right)_{exc} \rho_{e(h)} (\E) \nonumber \\
\left(\frac{dn_{e(h)}}{dt}\right)_{rec} &=& \int d\E \left(\frac{\partial f_{e(h)}}{\partial t}\right)_{rec} \rho_{e(h)} (\E) \nonumber \\
\left(\frac{dn_{e(h)}}{dt}\right)_{e(h)-ph} &=& \int d\E \left(\frac{\partial f_{e(h)}}{\partial t}\right)_{e(h)-ph} \rho_{e(h)} (\E) = 0 \nonumber \\
\left(\frac{dn_{e(h)}}{dt}\right)_{e(h)-e(h)} &=& \int d\E \left(\frac{\partial f_{e(h)}}{\partial t}\right)_{e-e} \rho_{e(h)} (\E) = 0. \nonumber \\
\end{eqnarray}
Here, the subscripts `$e$' and `$h$' stand for electrons and holes, respectively. At the steady-state, particle number conservation requires $\left(\frac{dn_{e(h)}}{dt}\right)_{exc}+\left(\frac{dn_{e(h)}}{dt}\right)_{rec} = 0$ for both electrons and holes separately. Both the electron (hole)-phonon and electron-electron (hole-hole) interactions are number conserving
The powers transferred between the different carriers in the conduction band are 
\begin{eqnarray}\label{SI-Eq-We}
W_{e-exc} &=& \int d\E ~\rho_e (\E) (\E-\E_c) \left(\frac{\partial f_e}{\partial t}\right)_{exc}, \nonumber \\
W_{e-rec} &=& -\int d\E \rho_e(\E) (\E-\E_c) \left(\frac{\partial f_e}{\partial t}\right)_{rec}, \nonumber \\
W_{e-ph} &=& \int d\E\rho_e(\E) (\E-\E_c) \left(\frac{\partial f_e}{\partial t}\right)_{e-ph}, \nonumber \\
W_{e-e} &=& \int d\E \rho_e(\E) (\E-\E_c) \left(\frac{\partial f_e}{\partial t}\right)_{e-e}  = 0 .
\end{eqnarray}
In the above, $W_{e-exc}$ is the power that electrons receive from the electromagnetic field due to photon absorption, and $W_{e-rec}$ is the power that electrons lose due to the recombination. Note that Eq. \eqref{SI-2-Eq:rec-elec} is negative and therefore, the definition of $W_{e-rec}$ in Eq.~\eqref{SI-Eq-We} makes $W_{e-rec}$ a positive quantity. The superscript `$(e)$' denotes that these powers correspond to the electrons in the conduction band, and the powers defined in~\eqref{SI-Eq-We} correspond to the rates of change of excess energy of electrons measured from the conduction band edge. Then, part of the net power stored in the electronic bath gets transferred to the phonon bath (i.e., the lattice) via electron-phonon scattering, and is given by $W_{e-ph}$. The electron-electron interaction, being an elastic scattering mechanism, does not induce any power transfer. The power balance equation corresponding to the electrons is given by
\begin{equation}\label{SI-Eq-Ue}
\frac{d \mathcal{U}_e}{dt} = c_e\frac{d T_e}{dt} = W_{e-exc} - W_{e-rec} - W_{e-ph},
\end{equation}
where $c_e$ is the electronic heat capacity.
In the non-equilibrium steady-state (NESS), we have $\frac{d \mathcal{U}_e}{dt} = 0$ which determines the steady-state electron temperature.

In analogy to the powers defined for the electrons in Eq. \eqref{SI-Eq-We}, for the holes we have
\begin{eqnarray}\label{SI-Eq-Wh}
W_{h-exc} &=& \int d\E ~\rho_h(\E) (\E_v - \E) \left(\frac{\partial f_h}{\partial t}\right)_{exc}, \nonumber \\
W_{h-rec} &=& -\int d\E~ \rho_h(\E) (\E_v - \E) \left(\frac{\partial f_h}{\partial t}\right)_{rec}, \nonumber \\
W_{h-ph} &=& \int d\E~\rho_h(\E) (\E_v - \E) \left(\frac{\partial f_h}{\partial t}\right)_{h-ph}, \nonumber \\
W_{h-h} &=& \int d\E~\rho_h(\E) (\E_v - \E) \left(\frac{\partial f_h}{\partial t}\right)_{h-h}  = 0,
\end{eqnarray}
calculated with respect to the valence band edge. The power balance equation corresponding to the holes is given by
\begin{equation}\label{SI-Eq-Uh}
\frac{d \mathcal{U}_h}{dt} = c_h \frac{d T_h}{dt} = W_{h-exc} - W_{h-rec} - W_{h-ph},
\end{equation}
where $c_h$ is the hole heat capacity.

Finally, we introduce a phenomenological description for the phonon temperature consisting of the total power transferred to the phonon sub-system (or the lattice) and the power transferred to the surrounding from the lattice. The power balance equation corresponding to the phonons is given by
\begin{eqnarray}\label{SI-Eq-Uph}
 \frac{d \mathcal{U}_{ph}}{dt} = c_{ph} \frac{d T_{ph}}{dt} = (W_{e-ph} + W_{h-ph})- G_{ph-env} (T_{ph} - T_{env}),
\end{eqnarray}
where $c_{ph}$ is the heat capacity of phonons and $G_{ph-env}$ coupling between the lattice and the environment. In the above equation, $\frac{d \mathcal{U}_{ph}}{dt}$ is the rate of change of energy of the phonon sub-system.


\section{Extraction of $T_e$ and $T_h$ from $W_{c-ph}^T$}\label{SI-2-Sec:delte_Weph}
The carrier-phonon power transfers at the thermal equilibrium, $W_{c-ph}^T$ are defined as
\begin{eqnarray}
W_{e-ph}^T &=& \int d\E (\E - \E_c) \left(\frac{\partial f_{e}^T}{\partial t} \right)_{e-ph} \rho_e(\E) \nonumber \\
W_{h-ph}^T &=& \int d\E (\E_v -\E) \left(\frac{\partial f_{h}^T}{\partial t} \right)_{h-ph} \rho_h(\E),
\end{eqnarray}
respectively, for electrons and holes, where $\left(\frac{\partial f_{e}^T}{\partial t} \right)_{e-ph}$ is given by Eq.~\eqref{SI-2-Eq:elec-ph-rate}  and $\left(\frac{\partial f_{h}^T}{\partial t} \right)_{h-ph}$ is given by Eq.~\eqref{SI-2-Eq:hole-ph-rate}.

To extract $T_e$ and $T_h$ from $W_{e-ph}^T$ and $W_{h-ph}^T$, respectively, we first obtain $W_{c-ph}^T$ as a function of $T_c - T_{ph}$ by taking $T_c$ as a variable, and then, we invert the $W_{c-ph}^T$ vs. $(T_c - T_{ph})$ relation to obtain $(T_c - T_{ph}) = \mathcal{F}(W_{c-ph}^T)$. 
Next, we find from $\mathcal{F}$ for which value of $(T_{e} - T_{ph})$ and $(T_{h} - T_{ph})$ we have $W_{e-ph}^{T} = W_{e-ph}$ and $W_{h-ph}^{T} = W_{h-ph}^{T}$, respectively, where $W_{e-ph}$ and $W_{h-ph}$ are obtained from our self-consistent calculation.
From $T_e - T_{ph}$ and  $T_h - T_{ph}$ we can further extract the rise in electron temperature, $\Delta T_e$, and the rise in hole temperature $\Delta T_h$ above that of the ambient. $\Delta T_e$ and $\Delta T_h$ are plotted as a function of $|E|^2$ in Fig.~3(a) of the main paper and exhibit a non-linear dependence on the intensity of the incident light.

The following method can be used to measure the temperature extracted from $W_{c-ph}$. At the steady-state we have $W_{c-exc} - W_{c-rec} - W_{c-ph}  = 0$, where we approximate $W_{c-ph} = G_{c-ph}(T_c, T_{ph}) (T_c - T_{ph})$ for both electrons and holes. Now, if we add a floating thermocouple (TC) for e.g., as considered in \onlinecite{dubi2011colloquium,cui2017study}, which measures the tunnelling of the electrons and holes, then, the energy flow equations for both electrons and holes get modified to
\begin{eqnarray}\label{SI-2-Eq:flow-eq}
 W_{c-exc} &-& W_{c-rec} - G_{c-ph} (T_c , T_{ph}) (T_c - T_{ph}) \nonumber \\ &-& G_{TC} (T_c -T_{TC}) = 0,
\end{eqnarray}
at the steady-state, where $c = e(h)$ for electrons (holes) and $G_{TC}$ is the coupling between the electronic subsystem of the SC and the TC. As a measuring probe we keep $T_{TC}$ a floating parameter and at the steady-state the flow equations reach at a fixed point where $T_{TC} = T_c$, giving us a unique value of carrier temperatures from the measurement.

\section{Steady-state Photoluminescence}\label{SI-2-Sec:ss-pl}
\begin{figure}
\includegraphics[scale=0.34]{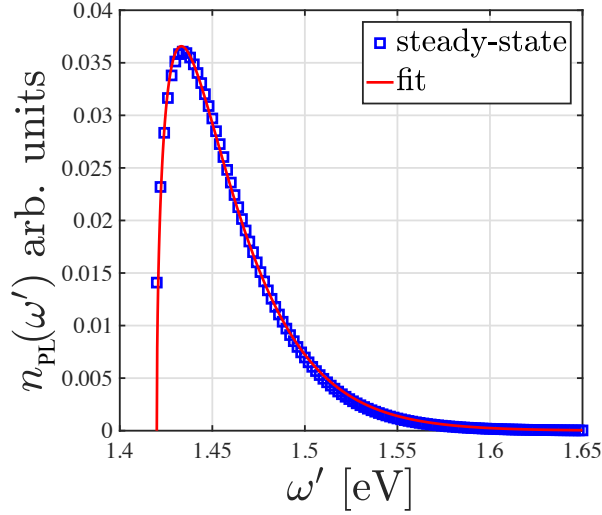}
\caption{\textbf{Steady-state photoluminescence at $|E|^2 = 40 ~ (V/m)^2$} Plot of the PL spectra as a function of the frequency of the emitted photons (in eV) where blue square data corresponds to the steady-state $n_{PL}(\omega')$ PL spectra, and the red solid line corresponds to a fit with $n_{PL}(\omega',T_c)$ where is the thermal PL spectra.
}
\label{SI-2-fig:ss-pl}
\end{figure}

Figure~\ref{SI-2-fig:ss-pl} plots the steady-state photoluminescence (PL) spectra for a specific intensity corresponding to $|E|^2 = 40~(V/m)^2$. The non-equilibrium PL spectrum $n_{PL}(\omega')$ is defined by integrating the right-hand-side of Eq.~\eqref{SI-2-Eq:rec_spem_rhoJ}. Here, we plot a normalized $n_{PL}(\omega')$ (in the sense that $\int_0^{\infty} d\omega' n_{PL}(\omega') = 1$). We normalize the PL spectra because we are interested only in its shape. We fit the steady-state PL spectrum with the thermal PL spectra defined by the right-hand-side of Eq.~\eqref{SI-2-Eq:n_rec} without the photonic density of states $\rho_{phot}$, whose functional form given by $n_{PL}(\omega', T_c) \propto \sqrt{\hbar \omega' -\E_g} e^{-\beta_c (\hbar \omega' -\E_g)}$ where $\beta_c = 1/k_B T_c$ \cite{photolum1, photoluminescence2}.

\paragraph*{Estimation of electron and hole from steady-state photoluminescence spectra:} 
The fitted data are plotted, after the due normalization, in Fig.~\ref{SI-2-fig:ss-pl} and provides us with a value of the carrier temperature $T_c$. The temperatures obtained from such a fitting for each values the intensity are plotted as a function of $|E|^2$ in Figure 3(b) of the main paper. For all energies $\E > \E_C$, $ \rho_{J, rec}^{c\rightarrow v}(\E - \hbar \omega^{\prime}, \E) = \rho_{J, rec}^{v\rightarrow c}( \E - \hbar \omega^{\prime} , \E)$, i.e., the joint density of states for an electron at $\E$ recombining with a hole at $\E - \hbar \omega'$ is the same as the joint density of states for a hole at $\E- \hbar \omega'$ recombining with an electron at $\E$, as seen from \eqref{SI-2-Eq:rec_spem_rhoJ} and \eqref{SI-2-Eq:rec_spem_rhoJ_hl}. This is also true for all energies $\E < \E_V$. Therefore, the PL spectrum is symmetric with respect to electrons and holes, making  the electron and hole temperatures obtained from the PL spectra the same.

\section{Ratio of power transfer for different photon energies}\label{Pl_eff_formalism}
\begin{figure}
\centering
\includegraphics[keepaspectratio=true,scale=0.255]{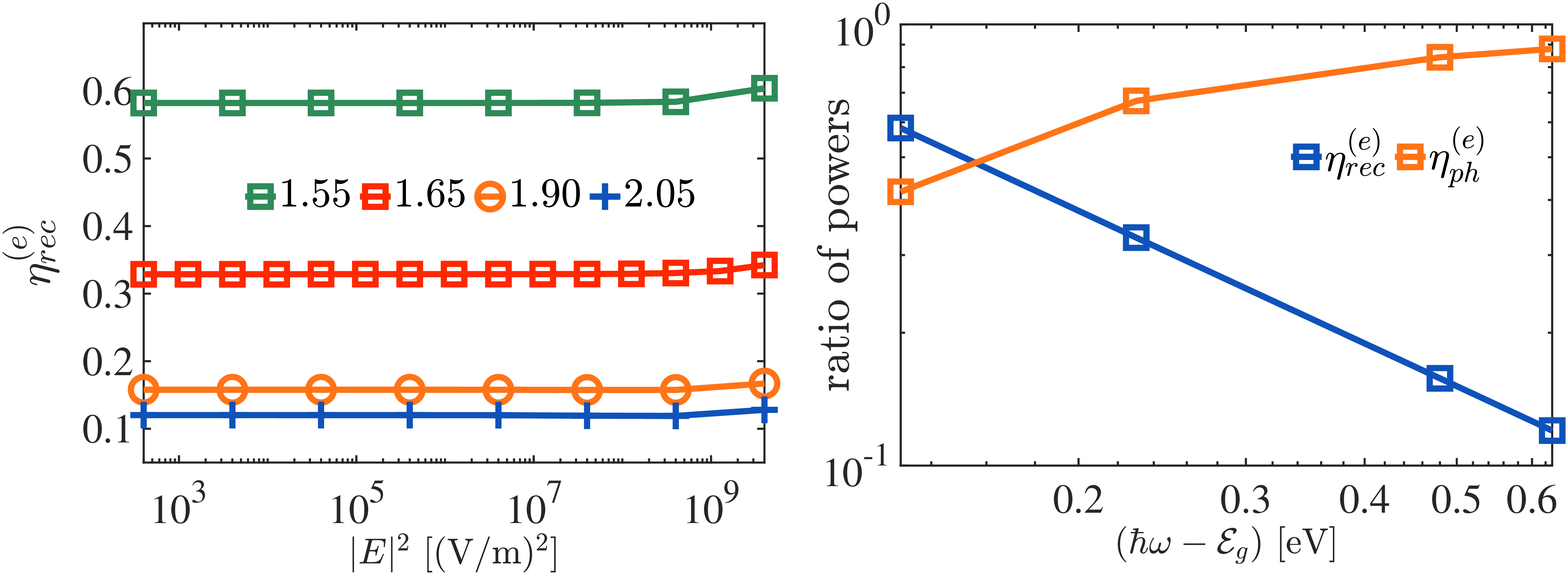}
\caption{Left Panel: $\eta_{rec}^{(e)}$ as a function of illumination intensities for four different values of photon energies $\hbar \omega$. Right panel: $\eta_{rec(ph)}^{(e)}$ vs. $(\hbar \omega - \E_{g})$ on a log-log scale at $|E|^2 = 4\times 10^{7}~(\frac{V}{m})^2$.}
\label{fig_eta_phteng}
\end{figure}

\textcolor{black}{In deriving the final form of Eq. \eqref{PL-eff} we define $\eta_{rec}^{(c)} = \frac{W_{c-rec}}{W_{c-exc}}$ [$c=e$ for electrons and $c=h$ for holes] as the ratio of the power dissipated from the electron/hole sub-systems through the recombination $W_{c-rec}$ to the power absorbed by the electron/hole sub-systems, $W_{c-exc}$. These ratios are plotted in Fig. \ref{eng_part} of the main text for $\hbar \omega = 1.65$ eV. Here in Fig. \ref{fig_eta_phteng} left panel, we plot $\eta_{rec}^{(e)}$ as a function of illumination intensities for four different values of photon energies $\hbar \omega$. Fig. \ref{fig_eta_phteng} right panel shows $\eta_{rec(ph)}^{(e)}$ vs. $(\hbar \omega - \E_{g})$ on a log-log scale at $|E|^2 = 4\times 10^{7}~(\frac{V}{m})^2$, where $\eta_{ph}$ is defined in Sec. \ref{sec:partition}.}
\textcolor{black}{Fig. \ref{fig_eta_phteng} right panel further shows that $\eta_{rec} \propto (\hbar \omega - \E_{g})^{-1}$ which is expected from the fact that $\eta_{rec}^{(c)} = W_{c-rec} / W_{c-exc}$ and $W_{c-exc} \propto (\hbar \omega - \E_{g})$. This further indicates that $W_{c-rec}$ is independent of the illumination photon energies. Moreover, $\eta_{ph}^{(c)}= \left( 1 -\frac{\text{const.}}{(\hbar \omega - \E_{g})} \right)$ due to $\eta_{rec}^{(c)}+\eta_{ph}^{(c)} = 1$ }

\bibliography{hot_electrons_in_semiconductors}

\end{document}